\documentstyle[11pt,epsf]{article}

\newcommand{ \Emax     }{ E_{\rm max} }
\newcommand{ \Ecrit    }{ E_{\rm crit} }
\newcommand{ \Etot     }{ E_{\rm tot} }
\newcommand{ \Einj     }{ E_{\rm inj} }
\newcommand{ \Edip     }{ E_{\rm dip} }
\newcommand{ \Ephzero  }{ E_{\ph_0} }
\newcommand{ \GeV      }{ {\rm GeV} }
\newcommand{ \LL       }{ {\cal L} }
\newcommand{ \MX       }{ M_{\scriptscriptstyle X} } 
\newcommand{ \tauX     }{ \tau_{\scriptscriptstyle X} }
\newcommand{ \etaX     }{ \eta_{\scriptscriptstyle X} }
\newcommand{ \nX       }{ n_{\scriptscriptstyle X} }
\newcommand{ \OmegaX   }{ \Omega_{\scriptscriptstyle X} }
\newcommand{ \qqbp     }{ q \overline{q}^{(')} }
\newcommand{ \ra       }{ \rightarrow }
\newcommand{ \ph       }{ \gamma }

\def\singleandabitspaced{\baselineskip=\normalbaselineskip\multiply
    \baselineskip by 115\divide\baselineskip by 100}
\def\singlespaced{\baselineskip=\normalbaselineskip}

\newcommand{ \centeron }[2]{{\setbox0=\hbox{#1}\setbox1=\hbox{#2}\ifdim
                             \wd1>\wd0\kern.5\wd1\kern-.5\wd0\fi \copy0
                             \kern-.5\wd0\kern-.5\wd1\copy1\ifdim\wd0>\wd1
                             \kern.5\wd0\kern-.5\wd1\fi}}
\newcommand{ \ltap }{\>\centeron{\raise.35ex\hbox{$<$}}
                     {\lower.65ex\hbox{$\sim$}}\>}
\newcommand{ \gtap }{\>\centeron{\raise.35ex\hbox{$>$}}
                     {\lower.65ex\hbox{$\sim$}}\>}
\newcommand{ \gsim }{\mathrel{\gtap}}
\newcommand{ \lsim }{\mathrel{\ltap}}

\newcommand{ \NPB    }[3]{Nucl. Phys. {\bf B#1}, #3 (#2)}
\newcommand{ \PLB    }[3]{Phys. Lett. B {\bf #1}, #3 (#2)}

\newcommand{ \PRD    }[3]{Phys. Rev. {\bf D#1}, #3 (#2)}

\newcommand{ \PREP   }[3]{Phys. Rept. {\bf #1}, #3 (#2)}

\newcommand{ \APJ    }[3]{Astrophys. J. {\bf #1}, #3 (#2)}
\newcommand{ \CPC    }[3]{Computer Physics Commun. {\bf #1}, #3 (#2)}

\begin{document}

\singlespaced

\begin{titlepage}

\begin{flushright}
hep-ph/9610468 \\
UCSD-TH-96-27 \\
October 1996
\end{flushright}
\vskip 1.0cm
\begin{center}
\vglue .06in
{\LARGE \bf Bounds on Long-Lived Relics from }
\vskip 0.1cm
{\LARGE \bf Diffuse Gamma Ray Observations}
\vskip 1.5cm
{\large Graham D. Kribs}\footnote{ {\small E-mail: kribs@umich.edu} } \\
\vskip 0.3cm
{\it Randall Physics Laboratory, University of Michigan, \\
     Ann Arbor, MI 48109-1120 } \\
\vskip 0.5cm
{\large I.Z. Rothstein}\footnote{ {\small E-mail: ira@yukawa.ucsd.edu} } \\
\vskip 0.3cm
{\it Dept. of Physics, UCSD, La Jolla, CA 92122 } \\
\vskip 2.0cm

\begin{abstract}
\indent

\singleandabitspaced

We place bounds on long-lived primordial relics using measurements 
of the diffuse gamma ray spectrum from EGRET and COMPTEL.  Bounds 
are derived for both radiative and hadronic decays with
stronger bounds applying for the latter decay mode.  We present an 
exclusion plot in the relic density--lifetime plane that
shows nontrivial dependence on the mass of the relic.
The violations of scaling with mass are a consequence of the different
possible scattering processes which  lead
to differing electromagnetic showering profiles.  The tightest bounds 
for shorter lifetimes come from COMPTEL observations of the low 
energy part of the spectrum, while for longer lifetimes the 
highest observable energy at EGRET gives the tightest bounds.
We discuss the implications of the bounds for dark matter candidates 
as well as relics that have a mass density substantially below the 
critical density.  These bounds can be utilized to eliminate models that
contain relics with lifetimes longer than $10^{-4}$ times the
age of the universe.

\end{abstract}

\end{center}
\end{titlepage}

\newpage
\setcounter{page}{2}
\singleandabitspaced

\section{Introduction}
\label{introduction-sec}
\indent

While the standard model of particle physics has passed all 
experimental tests to date, 
it lacks a particle candidate that could provide the dark 
matter of the universe as expected from astronomical observations.
Furthermore, our present understanding of structure formation 
seems to indicate that some fraction of the dark matter
should be ``cold'', so as to generate the proper power spectrum.
Such dark matter candidates are quite common in many
extensions of the standard model.  Indeed, many models
predict long lived relics that may or may not be dark matter candidates.
Long lifetimes for heavy relics, where by ``long'' we mean within
several orders of magnitude of the age of the universe 
may arise in many models which have symmetries that are only
broken at short distances.  Thus it is interesting to investigate 
the observational signatures of such long lived relics 
in an effort to rule out classes of models.

In this paper we study the signatures of particles with lifetimes
comparable to the age of the universe.
Such particles could play a role in solving the dark matter 
problem, but we will not confine our analysis to dark matter 
candidates.  The inclusion of long lived heavy ($\MX \gsim 50~\GeV$) 
particles necessitates an extension of the standard model.  
These relics could be 
technibaryons in technicolor models or the lightest supersymmetric
partner in an R-parity violating supersymmetric extension of the 
standard model.  The bounds found here
are model independent and depend on only three
parameters: the mass $\MX$, the lifetime $\tauX$
and the radiative or hadronic branching ratio times 
the relic density $B \cdot \etaX$, with $\etaX \equiv \nX/n_\ph$.
Given any model, it is possible to 
calculate the relic density using standard techniques leading to
bounds on couplings as well as masses.

Our exclusion bounds are derived by considering the direct 
observation of the gamma rays produced in the decay process. 
In general, the predicted observed spectrum will differ greatly 
from the decay spectrum due to redshifting and
scattering in the early universe.  The final spectrum can be 
compared to the EGRET and COMPTEL data leading to the exclusion 
plots presented here.

Previous investigations~\cite{barb,dod,bere3,ellis,gondolo} 
of the gamma ray spectra
produced by long lived relics have concentrated on radiative decays into
either photons or charged particles.  We consider both radiative
and hadronic decays\footnote{Some estimates for hadronic decays
were discussed in Ref.~\cite{barb}.}, including the effects of photon--photon 
scattering and $e^+e^-$ pair production.  There are reasons
to believe that hadronic decays are more compelling.
First, the hadronic branching ratio of the relics is expected 
to be of order one, unless there is a symmetry which forbids 
such decays\footnote{One would expect that the 
branching ratio into charged leptons may also be of order one, 
but then photons would only be generated if the lepton is 
energetic enough to shower.  See above and the footnote on 
p.\ 423 of Ref.~\cite{ellis}.  Bounds from measurements of the 
galactic positron flux were considered in Ref.~\cite{bere3}.}.
Radiative decays usually arise at the one loop level 
and so one would expect smaller branching ratios.  Second, 
hadronic decays produce more photons in the softer part of the 
spectrum due to fragmentation, which produces a large number
of pions that decay to two photons.  Thus, as opposed to
the case of radiative decay, we expect the non-scattered
spectra to have more photons at smaller energies
(i.e.\ $E_\ph \ll \MX$).  Since the diffuse photon 
background spectrum is well measured only up to $10$~GeV, 
we would naively expect our bounds to apply to larger mass relics
$\MX \ge {\cal O}(10^3)$~GeV for hadronic decays but not 
for radiative decays.  However, bounds derived from radiative
decays of large mass relics can be obtained if a reprocessing 
mechanism to lower the photon energy is active.
Such mechanisms begin operating once 
injected photon energies are above about $23$~GeV 
where  radiative decays can potentially compete with hadronic
decays and produce large numbers of photons at low energy
for masses larger than roughly $50$~GeV\@.  Thus, a
complete calculation is necessary to determine the bound 
for a given mode of decay, as we present here.

It should be noted that some of the bounds derived here 
will overlap those coming from structure formation arguments 
in the part of parameter space where the relic density 
is of the order of the critical density.
Radiative decay can lead to a radiation dominated epoch after
recombination that would drastically distort the
observed power spectrum.  Thus the shorter lifetimes considered
here could lead to such distortions, but we will not consider these effects.

\section{Electromagnetic Cascades in the Early Universe}
\label{em-cascades-sec}
\indent

\begin{figure}
\centering
\epsfxsize=3.5in
\hspace*{0in}
\epsffile{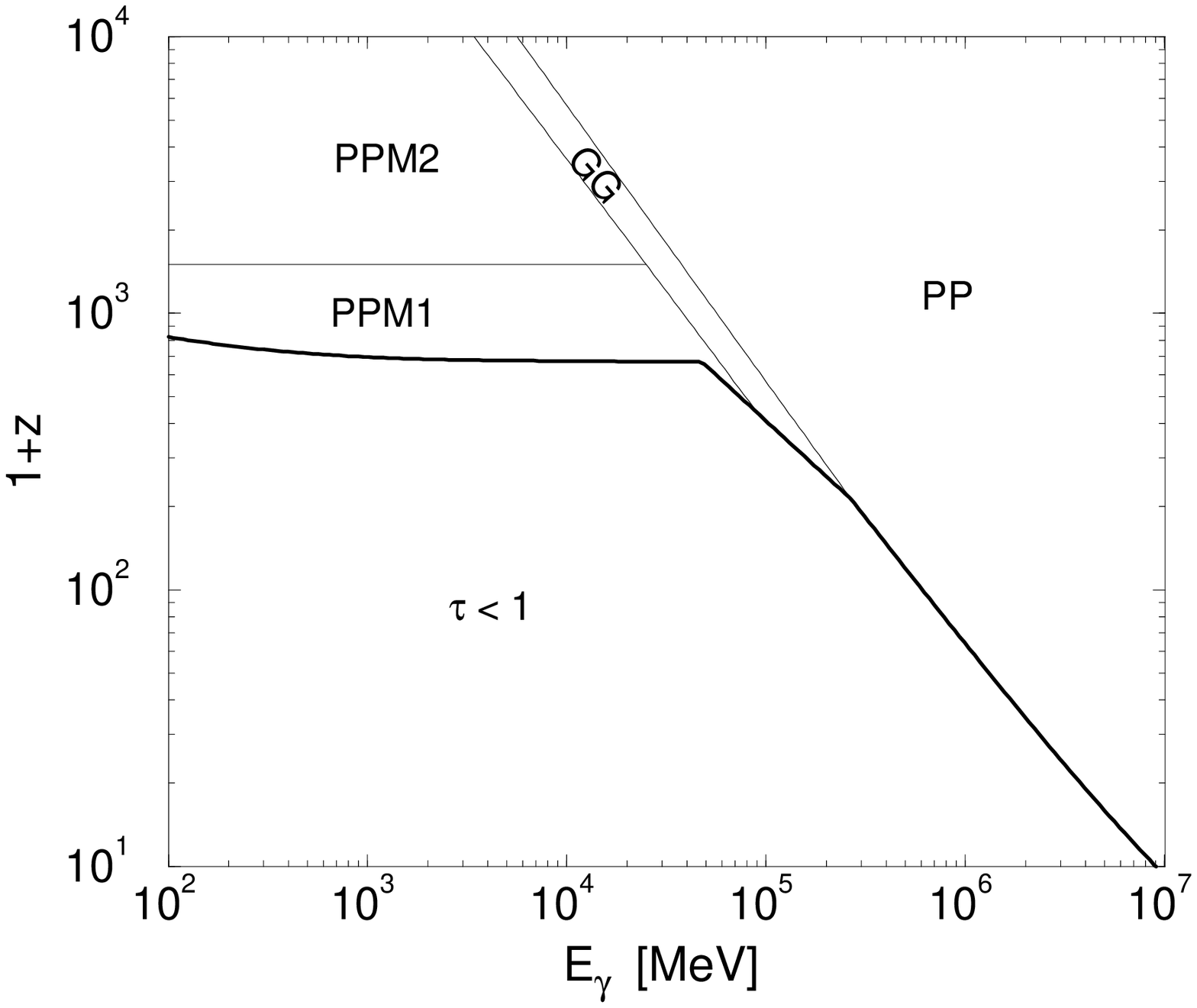}
\caption{Dominant scattering mechanisms for high energy photons
injected in the post-recombination era.  The region below the solid 
line has an optical depth $\tau < 1$, thus no scattering occurs.
The other regions are dominated by $e^+e^-$ pair production (PP), 
photon--photon scattering (GG), pair production in matter (PPM1)
and pair production in ionized matter (PPM2).}
\label{zE-region-fig}
\end{figure}

A high energy photon injected in the early universe will, 
in general, scatter.  Since there are many 
scattering processes possible, the nature of the 
scattering is strongly dependent on the redshift at which 
the photon is injected as well as its energy.  
Each process has a characteristic
optical depth  which determines its relevance to the evolution
of the photon.  The relevant processes were investigated in detail
by Zdziarski and Svensson in Ref.~\cite{zs}.  The processes include:
$e^+e^-$ pair production (PP), photon--photon scattering (GG), 
Compton scattering and pair production off of 
matter (PPM)\footnote{Pion production off of matter may 
also be of relevance for certain epochs, see Ref.~\cite{proth}.}.
Figure~\ref{zE-region-fig} (derived from~\cite{zs}) 
divides the redshift energy plane into regions labeled by 
the process that dominates in a particular part of the 
graph.  The photons always have injected energies, $E_\ph \gsim 100$~MeV,
which is obviously true of radiative decays of heavy relics 
and is also true of hadronic decays due to a cutoff
in the spectrum at $E_\ph = m_\pi/2$.  Therefore, the mechanisms 
for rescattering photons with energies below about $100$~MeV 
are irrelevant to our analysis.  Details of the spectra and 
cutoffs are described in the following sections.

Given the initial 
energy of the photon and the redshift at which it was injected, 
the progress of the injected photon can be tracked by moving 
horizontally across Fig.~\ref{zE-region-fig}.
For large enough photon energies, the time scale for scattering 
is short compared to the expansion 
rate of the universe, so we may neglect any vertical motion 
until we reach the region where Compton scattering dominates 
(not shown; to the left of the left edge of the graph) or where  the 
photon reaches the point where its optical depth drops below one. 
In the limit that the photon energy is much smaller than
the mass of the electron (Thompson limit), the cross section 
for Compton scattering is independent of energy.  When the 
photons reach the region of the plot dominated by Compton 
scattering in the Thompson limit, the photons will eventually 
be in kinetic equilibrium with the thermal bath leading to
a finite chemical potential for the photons.
The resulting distortion of the precisely measured microwave 
background leads to bounds on the injection of low-energy photons 
prior to recombination ($z \simeq 1300$), as discussed 
in Ref.~\cite{ellis}.  For photons with redshifts 
in the range $10^3 \lsim z \lsim 10^6$, the electromagnetic
cascades result in the production of $^3$He and D via the 
disintegration of $^4$He. Thus, there will be even more 
stringent bounds for this range of redshifts coming from 
limits on primordial abundances of light elements~\cite{lind,proth,ellis}. 
Here we concern ourselves with bounds coming from direct 
observations of gamma rays, and therefore we will be interested in 
photons whose life begins in the region dominated by pair 
production or photon--photon scattering 
after the epoch of recombination.  Such photons 
will eventually reach the region of the plot where the
optical depth drops below one and can be directly detected.

Pair production leads to a cascade, since the hot 
electron-positron pairs produced will inverse Compton 
scatter in the Klein-Nishina regime 
($\sigma \propto \ln(s)/s$), where they generate 
very energetic photons, which will in turn pair produce 
again.  This process will continue until the photon 
energies drop to the point where pair production off 
 the Wien tail of the black-body distribution is no 
longer efficient, or until we reach a regime where some other 
process (e.g.\ photon--photon scattering) begins to dominate.
Note that during this cascade process the mean free path 
for pair production is much smaller than the local Hubble 
length.  We may therefore calculate the cascade rate without 
regard to the effects of the expansion rate on the energies 
or occupation numbers.
 
For the region of energies and redshift relevant to our analysis,
photons will follow one of four paths.  If the photon 
is injected in the region where the optical depth is 
less than one, then the redshifted photon will free 
stream to the detector.  A photon injected in the region 
dominated by pair production will induce an electromagnetic 
cascade resulting in an ``escape photon spectrum'', that
was calculated in Ref.~\cite{sz,bere} and is presented in
the next section.  The spectrum terminates at $\Emax$, 
defined as the energy above which 
a photon will pair produce and degrade its energy.
The line dividing the PP and $\tau < 1$ regions in 
Fig.~\ref{zE-region-fig} is $\Emax$ as a function of redshift.
A photon born in the region dominated by pair production 
with a redshift in the range $300 < z < 700$ will have 
a slightly different fate\footnote{When $700 < z < 1300$ the 
photons undergo pair production off of matter, but the
resulting spectrum has not been calculated.  
Most of these photons will eventually reach kinetic 
equilibrium via Compton scattering.}.  In this regime, photon--photon
scattering becomes relevant~\cite{sz} since the energy of 
the photons are degraded and pass through the region GG 
in Fig~\ref{zE-region-fig}.
The width of this shaded region is given 
by~\cite{zs}
\begin{equation}
\frac{z_{\rm max}}{z_{\rm min}} \approx 
    \left( \frac{21}{\Omega_{0.1}h_{50}^2 T^{-3}_{2.7}}\right)^{1/3},
\end{equation}
where $\Omega_{0.1}=\Omega_b/0.1$, 
$h_{50}=H_0/(50~{\rm km}~{\rm s}^{-1}~{\rm Mpc}^{-1})$, 
and $T_{2.7}$ is the temperature of the microwave 
background in units of $2.7$~K\@.  The existence of this 
region has the effect of further distorting the photon
spectrum, as will be discussed in more detail below.
Finally, if a photon is injected directly into the region 
where photon--photon scattering dominates, it will lead to 
yet another spectrum of escape photons.

\section{Scattering processes}
\label{scattering-sec}
\indent

To quantify the photon scattering processes, 
we divide the region into three segments in redshift
$0 \le z \le 300$, $300 < z \le 700$, $700 < z$.  
This division in redshift, along with the forthcoming 
divisions in energy ($\Emax(z)$, $\Ecrit(z)$), provide
an approximation to Fig.~\ref{zE-region-fig} that we 
use throughout the following discussion of scattering
processes.  As anticipated,
we divide the region $0 \le z \le 300$ along the
line of optical depth $\tau = 1$ for pair production, 
defined by~\cite{sz}
\begin{equation}
\Emax = \frac{m_e^2}{30 T} \simeq \frac{36~{\rm TeV}}{1+z} 
\qquad (0 \le z < 300) .
\label{emax-lowz-eq}
\end{equation}
Photons injected
with redshifts in the region $0 \le z \le 300$ 
with energies $E_\ph < \Emax$ do not scatter,
while those with energies $E_\ph \ge \Emax$
pair produce and  generate a cascade.  The resulting spectrum 
is given by~\cite{zs,sz}
\begin{equation}
\frac{\LL(E_\ph)}{\Etot} = 
    \left\{ \begin{array}{ll}
            0.767 \, \Emax^{-0.5}E_\ph^{-1.5} 
                &  \mbox{ $0 \leq E_\ph < 0.04 \Emax$ } \\
            0.292 \, \Emax^{-0.2}E_\ph^{-1.8} 
                &  \mbox{ $0.04 \Emax \leq E_\ph < \Emax$ } \\
            0   &  \mbox{ $\Emax \leq E_\ph$ }.
            \end{array} \right.
\label{PP-lowz-eq}
\end{equation}
where $\LL$ is the number of photons per unit energy in
the spectrum and $\Emax$ is given by (\ref{emax-lowz-eq}).
The spectrum is normalized according to
$\int \LL(E_\ph) E_\ph dE_\ph = \Etot$,
where $\Etot$ is the fraction of energy in the injected 
spectrum above $\Emax$.

In the region $300 \le z < 700$, photon--photon scattering
dominates for photon energies above $\Ecrit$ and below $\Emax$, 
which is given by
\begin{equation}
\frac{\Ecrit}{\Emax} = \frac{z_{\rm min}}{z_{\rm max}} \sim \frac{1}{3}.
\label{ecrit-eq}
\end{equation}
$\Emax$ is now determined by equating the optical
depths for photon--photon scattering and pair production.
$\Emax$ is thus slightly larger~\cite{zs}
\begin{equation}
\Emax = \frac{m_e^2}{22 T} \simeq \frac{50~{\rm TeV}}{1+z}
\qquad (300 \le z < 700)
\label{emax-highz-eq}
\end{equation}
than in (\ref{emax-lowz-eq}).
Photons with $E_\ph < \Ecrit$ do not scatter, while
those with $\Ecrit \le E_\ph < \Emax$ photon--photon
scatter (with the background radiation),
and those with $\Emax \le E_\ph$ pair 
produce.  In the energy window $\Ecrit \le E_\ph < \Emax$,
each scattering of an energetic photon with a 
background photon results in the production of two photons 
which approximately share the energy of the injected photon.
The spectrum from these scattered photons takes the form~\cite{sz}
\begin{equation}
\frac{\LL(E_\ph)}{\Etot} = \frac{2.08}{\Ecrit^2 [1+(\Ecrit/\Einj)^3]^{1/3}
[1+(E_\ph/\Ecrit)^3]^{5/3}} \qquad (\Ecrit < \Einj < \Emax),
\label{GG-eq}
\end{equation}
where $\Einj$ is the energy of the assumed monoenergetic 
injected photons and $\Etot$ is the total integrated 
energy in the injected spectrum.  The spectrum is normalized
as in (\ref{PP-lowz-eq}), where we assume $\Ecrit/\Emax = 1/3$
to obtain the overall normalization constant.  A non-monoenergetic injected
spectrum can be treated by simply splitting injected spectra
into many small subregions between $\Ecrit$ and $\Emax$,
dividing the total integrated energy in the spectrum accordingly 
and using (\ref{GG-eq}).  The limiting behavior
of the resulting spectrum is proportional to a constant for
$E_\ph \ll \Ecrit$ and proportional to $E_\ph^{-5}$ for
$\Ecrit \ll E_\ph < \Einj$, thus  resulting in a startling spectral hump
near $\Ecrit$.  For photons injected with
energy $\Einj \ge \Emax$, pair production initially 
scatters the photons as in (\ref{PP-lowz-eq}),
but photons with energy below $\Emax$ can also rescatter by 
photon--photon scattering as above.  The scattered
spectrum can be approximated by~\cite{sz}
\begin{eqnarray}
\frac{\LL(E_\ph)}{\Etot} &=& 0.535 \left[ \frac{10}{\Ecrit^2 
      \left[ 1+ \left( \frac{E_\ph}{\Ecrit} \right)^3 \right]^{5/3}}
 \left[ 1 - \left( \frac{E_\ph}{\Emax} \right)^{0.2} \right] 
 \right. \nonumber \\
 & & \;\; \qquad \left. + \; \frac{1}{1+ \left( \frac{E_\ph}{\Ecrit} 
   \right)^3} \times
   \left\{ \begin{array}{ll}
            0.767 \, \Emax^{-0.5}E_\ph^{-1.5} 
                &  \mbox{ $0 \leq E_\ph < 0.04 \Emax$ } \\
            0.292 \, \Emax^{-0.2}E_\ph^{-1.8} 
                &  \mbox{ $0.04 \Emax \leq E_\ph \le \Emax$ }
            \end{array} \right. \right] 
\label{PP-highz-eq}
\end{eqnarray}
valid for $E_\ph \le \Emax$.  The limiting behavior of the spectrum
recovers the pair production result (\ref{PP-lowz-eq}) 
proportional to $E_\ph^{-1.5}$ for $E_\ph < 0.04 \Emax$, while 
the first term in (\ref{PP-highz-eq}) dominates between
$0.07 \Emax \lsim E_\ph \lsim 0.86 \Emax$ and the second
term dominates for $E_\ph \gsim 0.86 \Emax$ leading to a spectrum 
proportional to $E_\ph^{-4.8}$.

\begin{figure}
\centering
\epsfxsize=3.5in
\hspace*{0in}
\epsffile{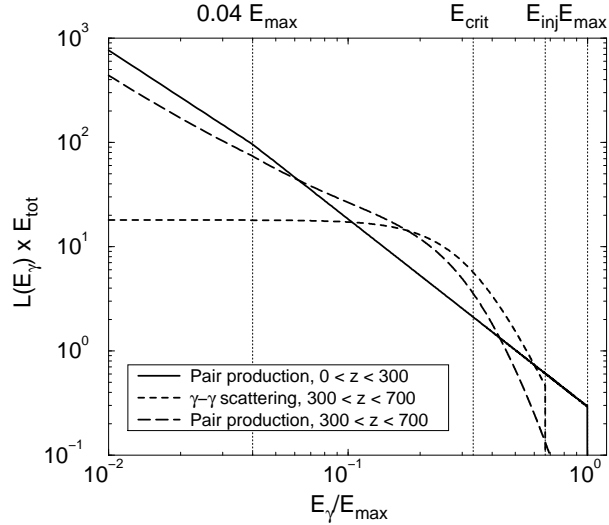}
\caption{Scattered spectra from pair production at low $z$,
photon--photon scattering, and pair production at high $z$.
Each spectra has a total integrated energy of $\Etot$, with
$\Emax$ set to unity.  See the text for details.}
\label{scatter-fig}
\end{figure}

In Fig.~\ref{scatter-fig} we show the spectra for pair production
in the low and high $z$ regimes and the spectra for 
photon--photon scattering, assuming a total integrated energy
of $\Etot$ in each case.  The energy is normalized 
with $\Emax = 1$, $\Ecrit = \Emax/3$ and $\Einj = 2\Emax/3$,
the latter being an arbitrary choice (within the allowed range 
$\Emax/3 < \Einj < \Emax$) for illustration.

\section{Redshifting}
\label{redshifting-sec}
\indent

Given a photon spectrum after decay (from direct and/or reprocessed
photons) $\LL(E_\ph)$, the spectrum we see today is
simply an integral over all redshifts convoluted with 
exponential decay rate,
\begin{equation}
\frac{dJ}{d \Ephzero} \; = \; \frac{3}{8\pi}\frac{t_0}{\tauX}
   \int^{z_0}_0\frac{dz}{(1+z)^{3/2}} \LL(E_\ph) 
   \left( \frac{n_\ph(t_0)}{n_\ph}\nX \right) 
   \exp \left[ -\frac{t_0}{\tauX}(1+z)^{-3/2} \right],
\label{redshift-eq}
\end{equation}
where $\frac{dJ}{d \Ephzero}$ is the flux of photons,
$\Ephzero$ is the present-day photon energy, 
$t_0$ is the age of the universe, $\tauX$ is the lifetime 
of the relic, $n_\ph(t_0)$ is the present-day density of photons, 
and $n_\ph$ and $\nX$ are respectively the densities 
of the photons and relics at decoupling
(we use $\Omega = 1$ from here on).  From the previous 
section, we use $z_0 = 700$ as the upper limit in redshift 
since high energy photons injected above this redshift
will pair produce off of matter and eventually Compton scatter
thus avoiding direct detection. The photon spectrum today in 
(\ref{redshift-eq}) can be written suggestively as
\begin{equation}
\frac{dJ}{d \Ephzero} \; = \; B_\ph \etaX \left\{ 1.5 \times 10^{9} \> 
   \frac{t_0}{\tauX} \int^{700}_0\frac{dz}{(1+z)^{3/2}} 
   \frac{\LL (E_\ph)}{B_\ph} 
   \exp \left[ -\frac{t_0}{\tauX}(1+z)^{-3/2} \right] \right\} 
   \quad ({\rm cm}^2 {\rm ~s~sr~MeV})^{-1},
\label{redshift-rewrite-eq}
\end{equation}
where $\LL(E_\ph)$ is in [GeV]$^{-1}$.  It is clear from
(\ref{redshift-rewrite-eq}) that an upper limit on the present-day 
flux of photons can be translated into an upper limit on `the ratio 
of relic to photon number density' $\etaX$ (times the relic branching 
ratio to photons $B_\ph$) at $z=700$.

 If the lifetime is short relative to the age 
of the universe, then most of the  photons would be reprocessed 
at or before recombination, and would not reach the detector.  
Thus, to ensure that an appreciable number of
  photons can be observed today, 
we require
\begin{equation}
\tauX/t_0 \gsim (1+z_0)^{-3/2} \approx 5 \times 10^{-5}.
\label{shortest-lifetime-eq}
\end{equation}
Furthermore, we have specified that the bound we obtain is the
relic density at $z_0$.  This is roughly equivalent to the
relic density at decoupling if the lifetime is longer than
(\ref{shortest-lifetime-eq}), so that the density does not
change appreciably between decoupling and $z_0$.

\section{Diffuse Photon Background}
\label{measurements-sec}
\indent

To establish bounds on the relic density, we use the recent
bounds on the extra-galactic diffuse gamma ray background from
the EGRET~\cite{EGRET} and COMPTEL~\cite{COMPTEL} instruments
aboard the Compton Gamma Ray Observatory.  Both instruments
find that the diffuse photon flux obeys a power law
\begin{equation}
\frac{dJ}{d\Ephzero} \; = \> 
   k \left( \frac{\Ephzero}{1 \; {\rm MeV}} \right)^{-\alpha} 
   \qquad ({\rm cm}^2 {\rm ~s~sr~MeV})^{-1} ,
\end{equation}
roughly consistent with (but more sensitive than) measurements done 
by the \mbox{SAS-2} experiment many years ago~\cite{SAS}.  
EGRET fit to a power law for photons
in its observable range $30 < \Ephzero < 10^{4}$~MeV and found
$k = 2.26 \times 10^{-3}$ and $\alpha = 2.07$.
COMPTEL also found a reasonable fit with $\alpha \simeq 2.3$
(although they did not explicitly give fit parameters
with errors).  We take the EGRET fit to be valid down to
$\Ephzero = 30$~MeV, then we estimated a best fit power law
for the COMPTEL data that is continuous through
$\Ephzero = 30$~MeV\@.  We obtained $\alpha = 2.38$ with
$k = 6.4 \times 10^{-3}$ which fits the COMPTEL data
quite well down to $\Ephzero = 0.8$~MeV (the lowest energy
reported), and also fits other data~\cite{otherdata} below
the sensitivity of COMPTEL to roughly $\Ephzero = 0.1$~MeV\@.
We also note here that the infamous ``MeV bump'' discussed
in Ref.~\cite{ellis} has disappeared.

%
%

\section{Bounds from Radiative Decays}
\label{rad-decays-sec}
\indent

\subsection{Preamble and Previous Results}
\indent

Bounds on relics with radiative decays from diffuse background 
measurements have been considered previously in 
Refs.~\cite{ellis,dod,gondolo}. Our analysis  differs in several
ways.  Many of the bounds derived here were found using 
the new EGRET data which allows us to look at higher energy 
gamma rays.  Furthermore, our analysis
of the showering profiles differs from those given in~\cite{ellis,gondolo}.
The authors of \cite{ellis,gondolo} determined $\Emax$ by equating the
Compton scattering cross section with that from pair production, 
leading to a much lower value of $\Emax$ than what was used 
in this paper. 
This lower value of $\Emax$ leads to showering at lower values of the
relic mass and thus the bounds found in~\cite{ellis,gondolo} have a 
different mass dependence than found here. 
As discussed in the previous section, 
Compton scattering does not become 
important until much lower energies~\cite{sz} and does not play
a role in the determination of $\Emax$. 
We also included a more complete analysis of the 
spectral distortion due to photon--photon scattering
than was considered in~\cite{ellis}, though the effects on the
bounds are minor. 

In what follows, we assume 2-body decays $X \ra \ph\ph$ 
with a branching ratio $B_\ph$, giving a (non-scattered) input
spectrum
\begin{equation}
\frac{\LL(E_\ph)}{\Einj} = \frac{B_\ph}{\Einj} \delta( E_\ph - \Einj ),
\label{radiative-noscat-eq}
\end{equation}
where we use $\Einj = \MX/2$ to represent the injected energy 
per decay, with a total energy in the input spectrum of
$\Etot = 2 B_\ph \Einj$ (the factor of $2$ due to two photons
in the final state).  The bounds we find for 2-body decays can
be applied approximately for 3-body decays by setting 
$\Einj = \langle E \rangle = \MX/3$.  In addition, 2-body 
decays to single photons $X \ra Y\ph$ can be similarly
constrained by scaling up the bound on $\etaX$ by a factor
of two.

\subsection{No scattering}
\indent

\begin{figure}
\centering
\epsfxsize=3.5in
\hspace*{0in}
\epsffile{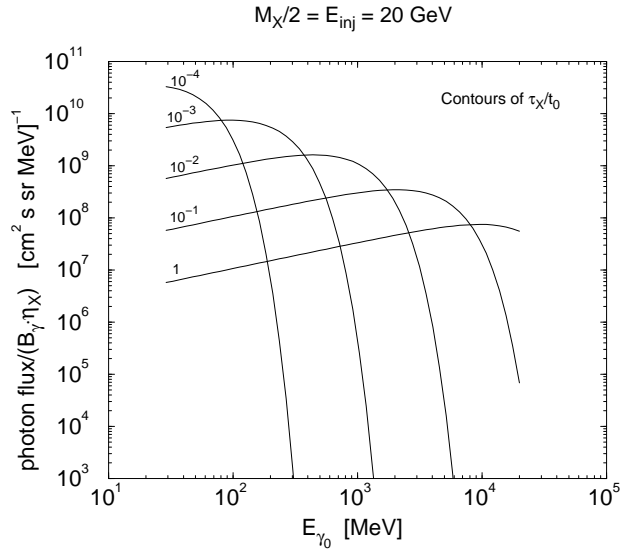}
\caption{Photon flux as a function of photon energy ($z = 0$) 
from a non-scattered 2-body decay (see (\ref{radiative-noscat-eq})) 
with $\MX/2 = \Einj = 20$~GeV\@.
The spectra scale linearly with the radiative branching ratio $B_\ph$ 
and the relic density $\etaX$,
thus an observational limit on the photon flux can be 
translated into a limit on $B_\ph \etaX$.  
A sample of relic lifetimes $\tauX/t_0$ were chosen 
and plotted as separate contours on the graph.
}
\label{2-body-20-fig}
\end{figure}

In the special case that the relic decays with no 
scattering (so that (\ref{radiative-noscat-eq}) {\em is}\/
the input spectrum), the present-day photon spectrum can be calculated
exactly from (\ref{redshift-eq}) yielding
\begin{equation}
\frac{dJ}{d\Ephzero} = \frac{3}{8\pi}B_\ph \etaX n_{\ph}(t_0)
   \frac{t_0}{\tauX}\frac{E^{1/2}_{\ph_0}}{\Einj^{3/2}}
   \exp\left[-\frac{t_0}{\tauX}\left(\frac{\Ephzero}
   {\Einj}\right)^{3/2}\right],
\end{equation}
for $\Einj/(1+z_0) < \Ephzero < \Einj$.  The photon flux
rises proportional to $\Ephzero^{1/2}$ up to 
roughly $\Ephzero \approx \Einj ( \tauX/t_0 )^{2/3}$ and then drops
exponentially for higher $\Ephzero$ up to $\Einj$.  This can be seen in 
Fig.~\ref{2-body-20-fig} where sample spectra are shown for 
$\Einj = 20$~GeV with relic
lifetimes in the range $10^{-4} \le \tauX/t_0 \le 1$.
The lower bound $\Ephzero = \Einj/(1+z_0) \simeq 28.5$~MeV
is clearly visible as a cutoff in the spectrum, as is the upper 
bound from the photon injection energy $\Ephzero = \Einj = 20$~GeV\@.
Notice that for short lifetimes $\tauX/t_0 \lsim 10^{-4}$,
the exponential suppression completely dominates the final
spectra for all $\Ephzero$.


\subsection{Numerical results, with scattering}
\indent 

\begin{figure}
\centerline{
\hfill
\epsfxsize=0.60\textwidth
\epsffile{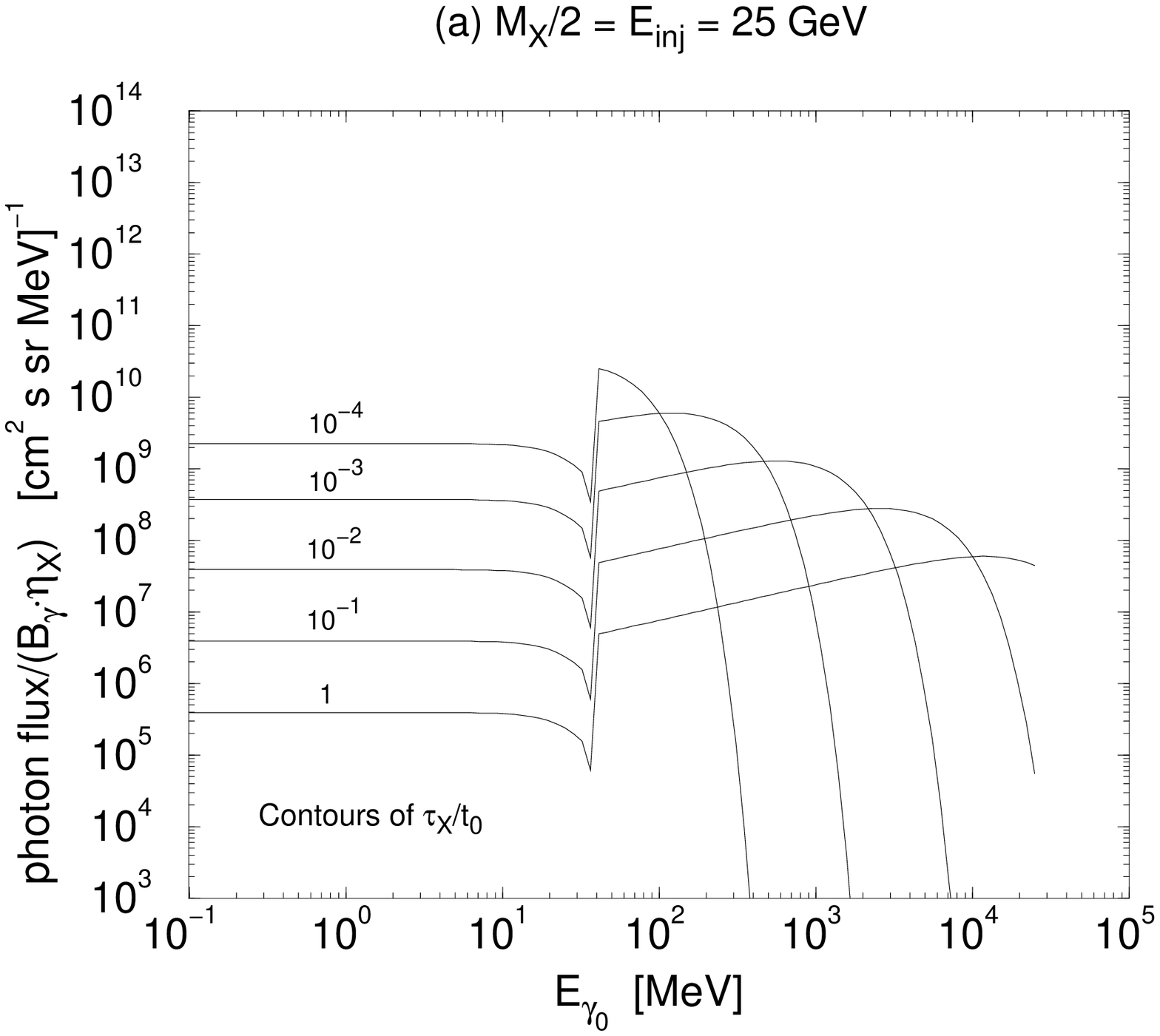}
\hfill
\epsfxsize=0.60\textwidth
\epsffile{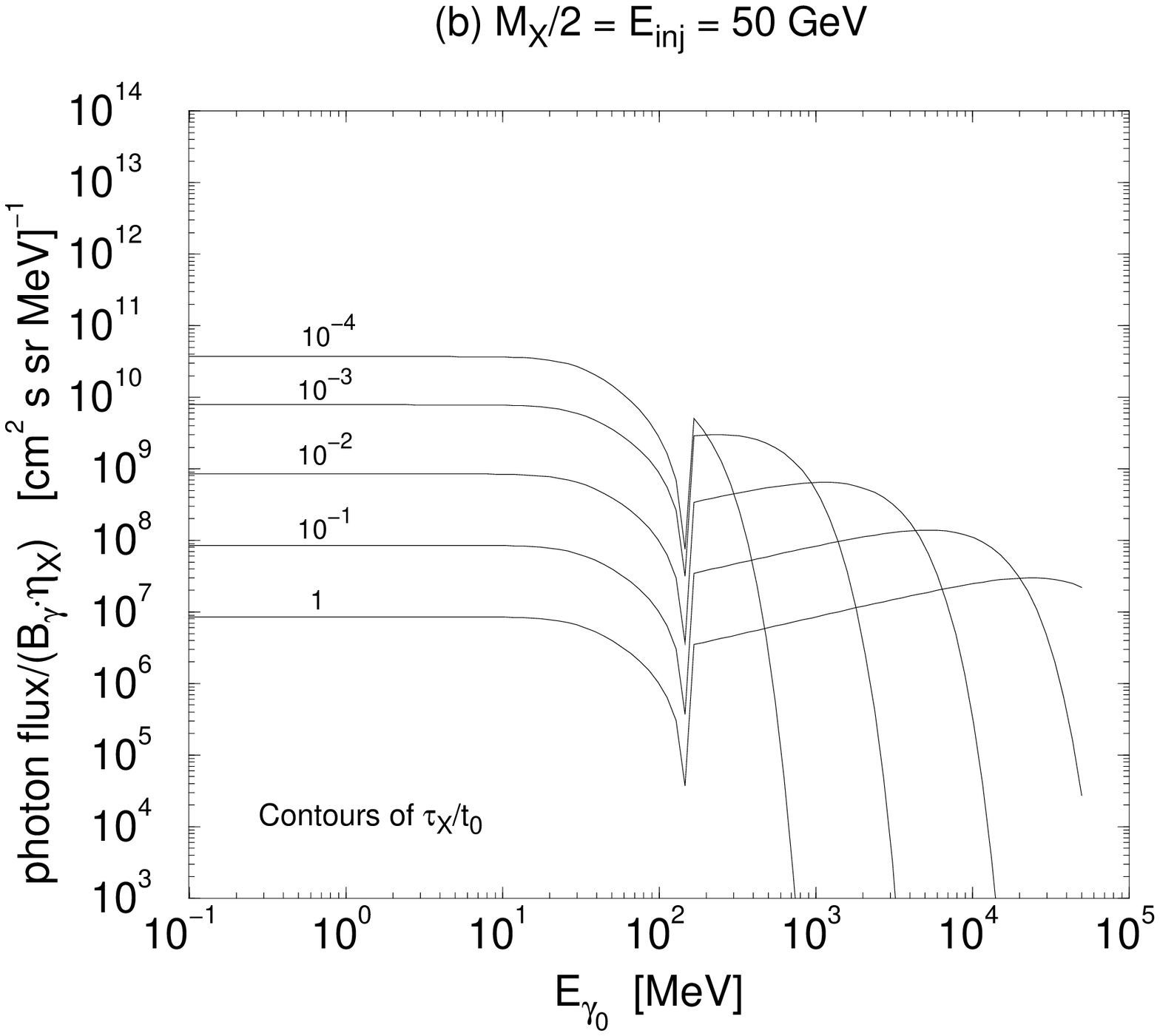}
\hfill }
\centerline{
\hfill
\epsfxsize=0.60\textwidth
\epsffile{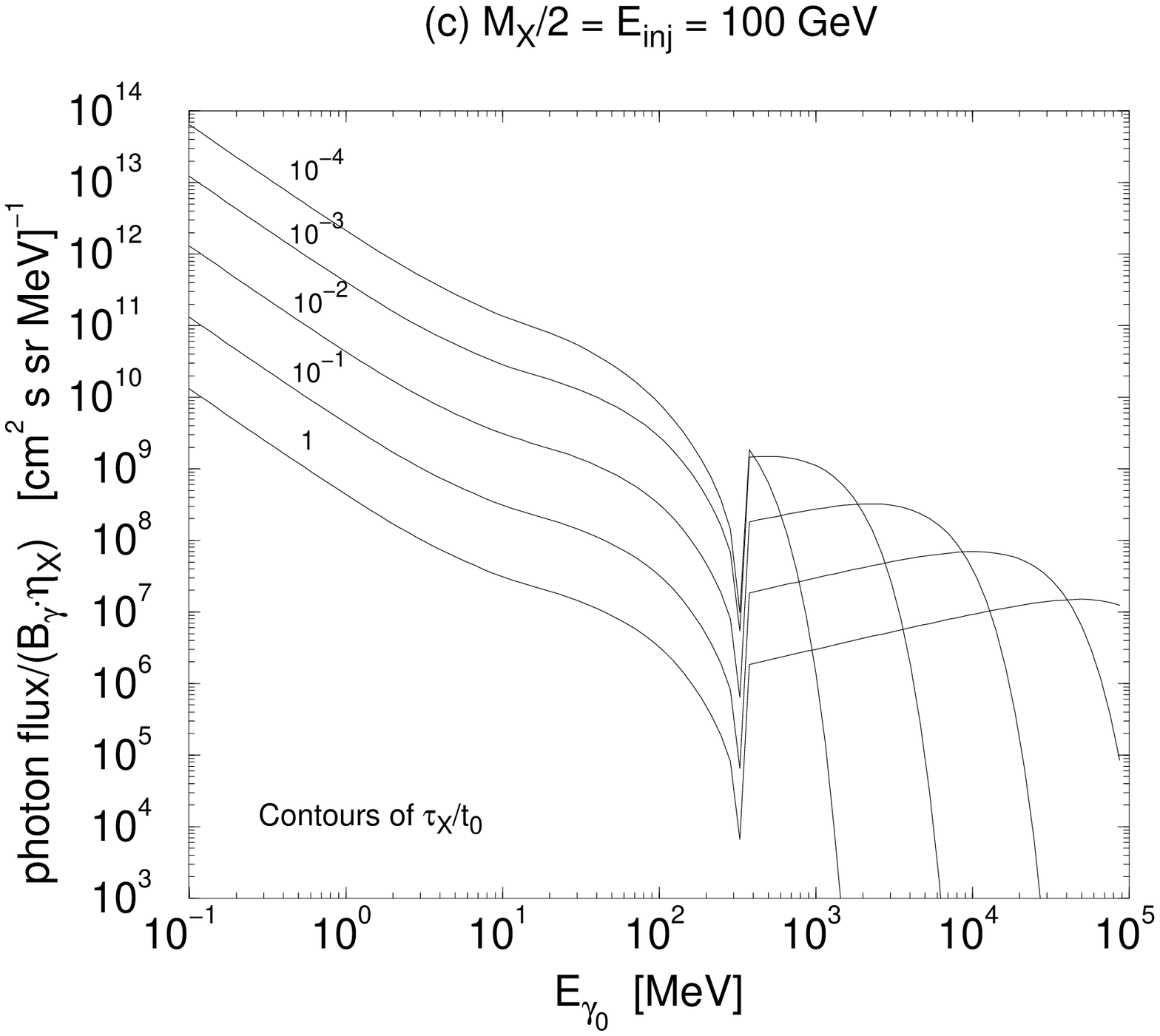}
\hfill
\epsfxsize=0.60\textwidth
\epsffile{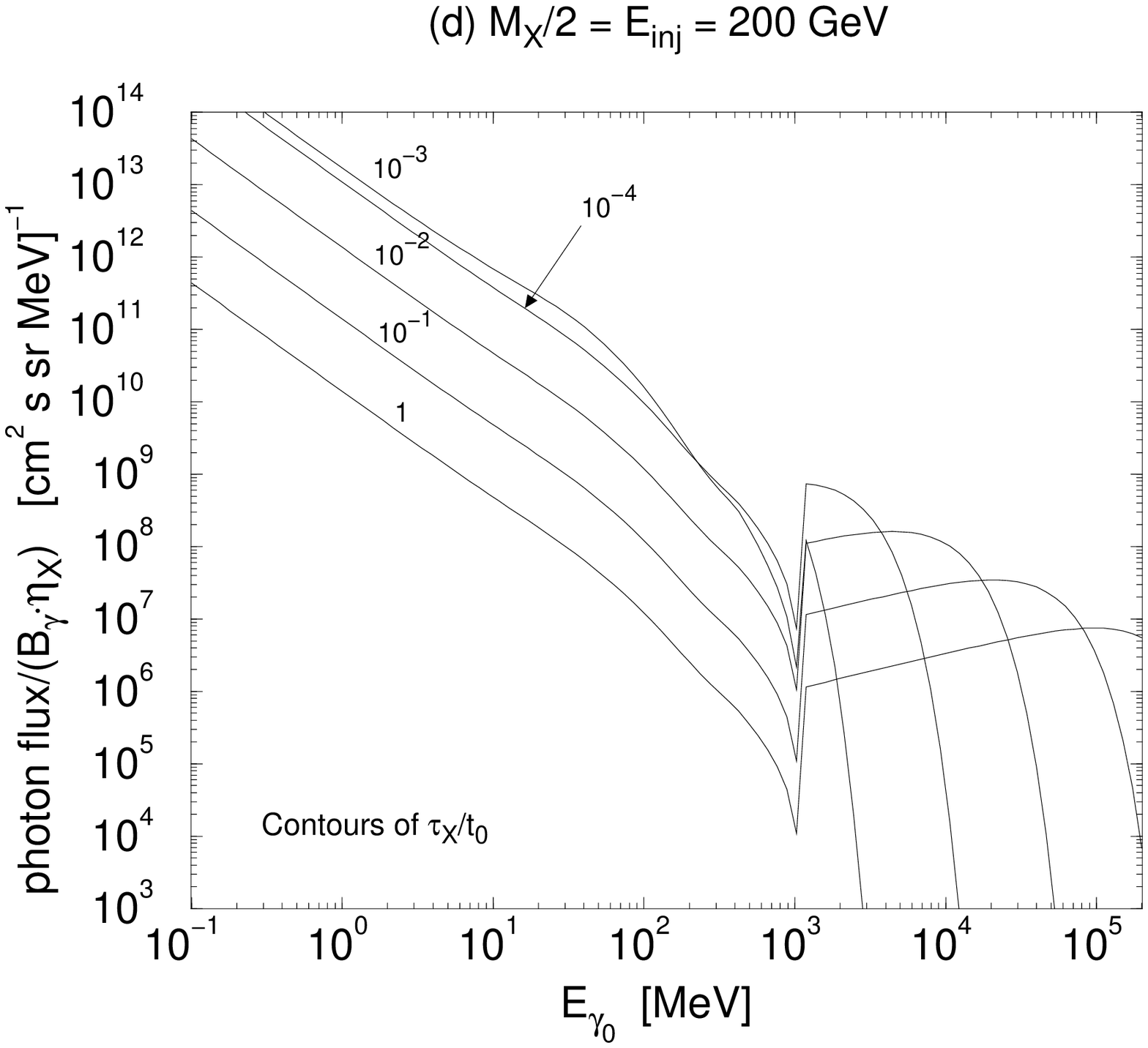}
\hfill }
\caption{Redshifted photon spectra for a relic with
mass $\MX/2 = \Einj = 25$, $50$, $100$, $200$~GeV, 
a 2-body decay $X \ra \ph\ph$ with branching ratio $B_\ph$,
and a relic density $\etaX$.
A sample of relic lifetimes $\tauX/t_0$ were chosen 
and plotted as separate contours on the graph.
}
\label{2-body-25-200-fig}
\end{figure}

The sample spectra in Fig.~\ref{2-body-20-fig} illustrate
the effect of redshifting, but the more general case
with scattering is what is  of interest to determine relic
density times branching ratio bounds.
In Table~\ref{scattering-onset-table} we list the relevant 
scattering mechanisms with their redshift and injected 
energy dependence.  
For a given injected photon energy, we expect a dip 
in the spectrum due to the transition from 
scattered to unscattered photons as $\Einj$ is increased.
The dip is located at
\begin{equation}
\Edip \; = \; \left\{ \begin{array}{ccl}
    \frac{\Einj^2}{\Ecrit(z=0)} & & (\Einj < 55~{\rm GeV}) \\
    \frac{\Einj}{301}           & & (55 < \Einj < 122~{\rm GeV}) \\
    \frac{\Einj^2}{\Emax(z=0)}  & & (\Einj > 122~{\rm GeV})
    \end{array}
    \right. ,
\label{edip-eq}
\end{equation}
where 
\begin{equation}
\Ecrit(z=0) = \frac{m_e^2}{3 \cdot 22 T_0} \approx 17~{\rm TeV}
\end{equation}
and
\begin{equation}
\Emax(z=0) = \frac{m_e^2}{30 T_0} \approx 36~{\rm TeV} .
\end{equation}
The dependence on $\Einj$ comes from the
fact that at larger energies it is possible to scatter at 
smaller redshifts.   This is readily seen in 
the $\Einj = 25, 50$~GeV figures of Fig.~\ref{2-body-25-200-fig}. 
In Table~\ref{2-body-crit-table} we evaluate
(\ref{edip-eq}) for the injected energies that we have considered.
The photons with energies smaller than 
(or to the left of) $\Edip$ are those that were reprocessed 
by scattering, and thus decayed at an earlier redshift 
then those with energy larger than (or to the right of) $\Edip$, that were 
unprocessed.  As we further increase $\Einj$, pair production 
turns on as seen in the $\Einj = 100$, $200$~GeV figures of 
Fig.~\ref{2-body-25-200-fig}.  In these figures we see the 
power law behavior expected at low energies coming from those 
photons produced in the pair production cascade.

Notice in particular that $\Einj = 100$~GeV
displays a scattered spectra that is consistent 
with (\ref{PP-highz-eq}), whereas the scattered 
spectra for $\Einj \ge 200$~GeV is consistent with 
(\ref{PP-lowz-eq}) (see Fig.~\ref{scatter-fig}).  
The reason for this difference is that the threshold for
pair production at low $z (< 300)$ is crossed once 
$\Einj \ge 200$~GeV, and consequently those scattered 
photons dominated the final spectra.  It is also important
to notice the critical injected energy $\Einj = 600$~GeV
corresponds to an $\Edip = 10$~GeV 
in Table~\ref{2-body-crit-table}.  For injection 
energies above this value, the only detectable photons will 
be those which undergo photon--photon scattering.

\begin{table}
\renewcommand{\baselinestretch}{1.2}\small\normalsize
\begin{center}
\begin{tabular}{rclcl} \hline\hline
\multicolumn{3}{c}{$\Einj$} & & Scattering mechanism(s) \\ \hline
      & $<$   & $23$~GeV    & & no scattering  \\
$23$  & $\ra$ & $55$~GeV    & & photon--photon scattering
                                ($1+z > \frac{17~{\rm TeV}}{\Einj}$) \\
$55$  & $\ra$ & $71$~GeV    & & photon--photon scattering
                                ($1+z > 301$) \\
$71$  & $\ra$ & $167$~GeV   & & photon--photon scattering
                                ($301 < 1+z < \frac{50~{\rm TeV}}{\Einj}$) \\
$71$  & $\ra$ & $167$~GeV   & & pair production 
                                ($1+z > \frac{50~{\rm TeV}}{\Einj}$) \\
$122$ & $\ra$ & $167$~GeV   & & pair production 
                                ($\frac{36~{\rm TeV}}{\Einj} < 1+z < 301$) \\
$167$ & $\ra$ & $3.7 \times 10^{4}$~GeV   & & pair production 
                                ($1+z > \frac{36~{\rm TeV}}{\Einj}$) \\
      & $>$   & $3.7 \times 10^{4}$~GeV   & & pair production (all $z$) \\ 
  \hline\hline
\end{tabular}
\end{center}
\renewcommand{\baselinestretch}{1.0}\small\normalsize
\caption{Scattering mechanisms for different injected
photon energies $\Einj$ and redshifts $z$.
}
\label{scattering-onset-table}
\end{table}

\begin{table}
\renewcommand{\baselinestretch}{1.2}\small\normalsize
\begin{center}
\begin{tabular}{cc} \hline\hline
$\Einj$ (GeV) & $\Edip$ (GeV)  \\ \hline
$25$        & $0.037$  \\
$50$        & $0.15$   \\
$100$       & $0.33$   \\
$200$       & $1.1$    \\
$400$       & $4.4$    \\
\mbox{\boldmath $600$}   &  \mbox{\boldmath $10$}   \\
$800$       & $17$     \\
$1600$      & $70$     \\
$3200$      & $280$    \\ 
$6400$      & $1100$   \\ \hline\hline
\end{tabular}
\end{center}
\renewcommand{\baselinestretch}{1.0}\small\normalsize
\caption{The location of the dip in redshifted spectra $\Edip$, 
where above (below) this value the photons originated from
unscattered (scattered) injected photons (see (\ref{edip-eq})).
The entry in bold $\Edip = 10$~GeV, corresponding to an 
injected energy $\Einj = 600$~GeV is a critical point
where injected energies above this value can only be 
detected today through scattered photons.
}
\label{2-body-crit-table}
\end{table}

\begin{figure}
\centerline{
\hfill
\epsfxsize=0.60\textwidth
\epsffile{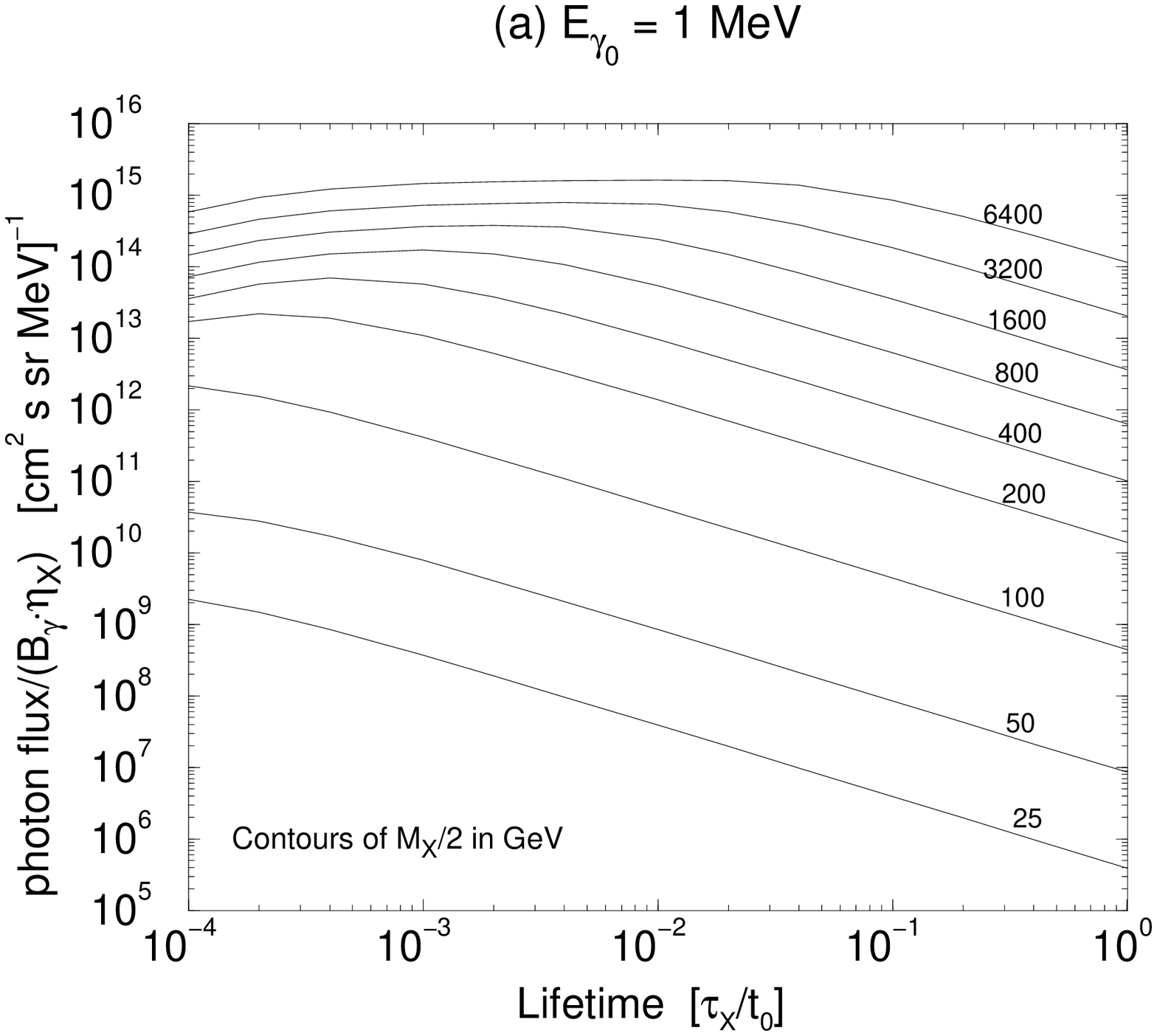}
\hfill
\epsfxsize=0.60\textwidth
\epsffile{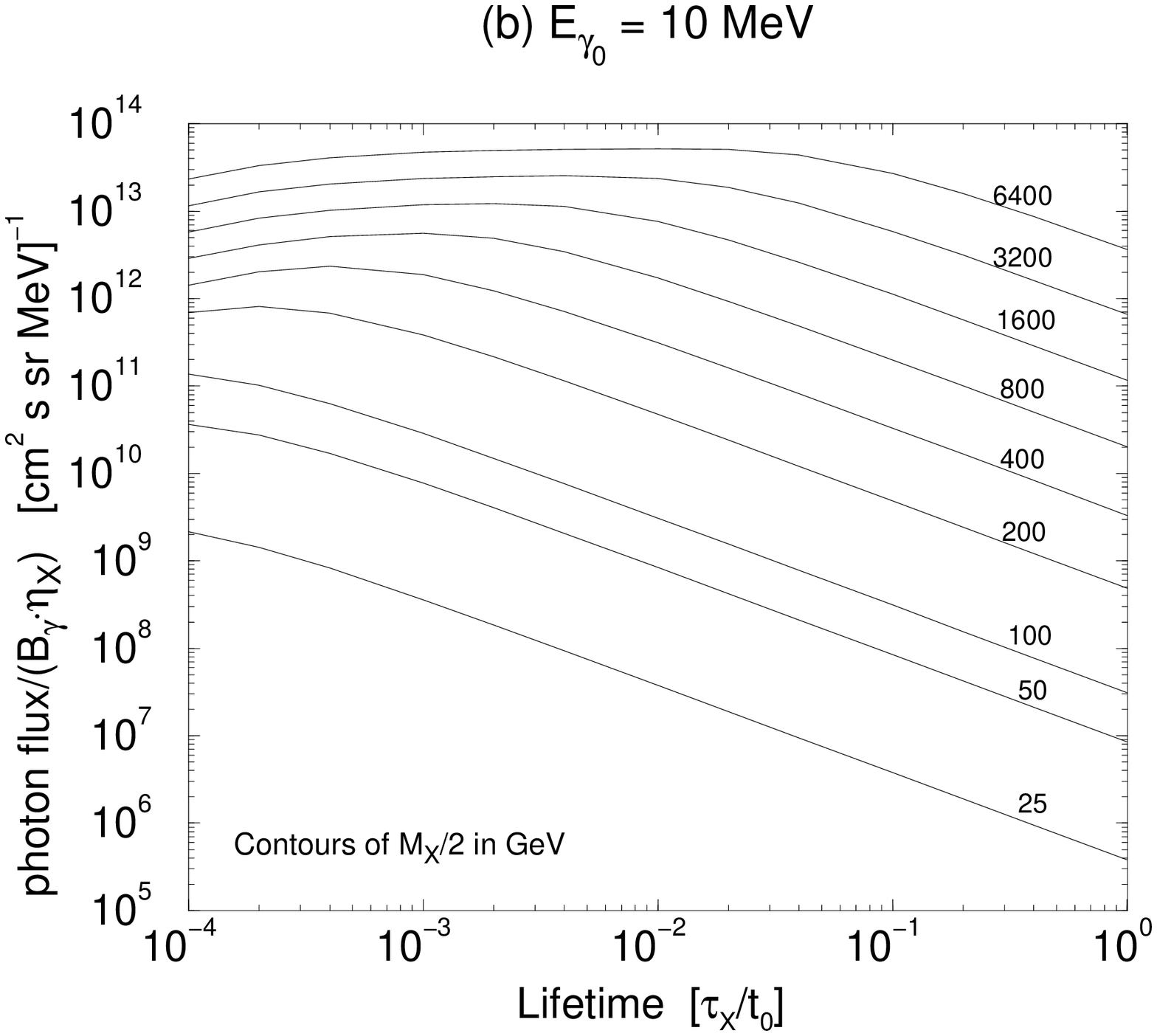}
\hfill }
\centerline{
\hfill
\epsfxsize=0.60\textwidth
\epsffile{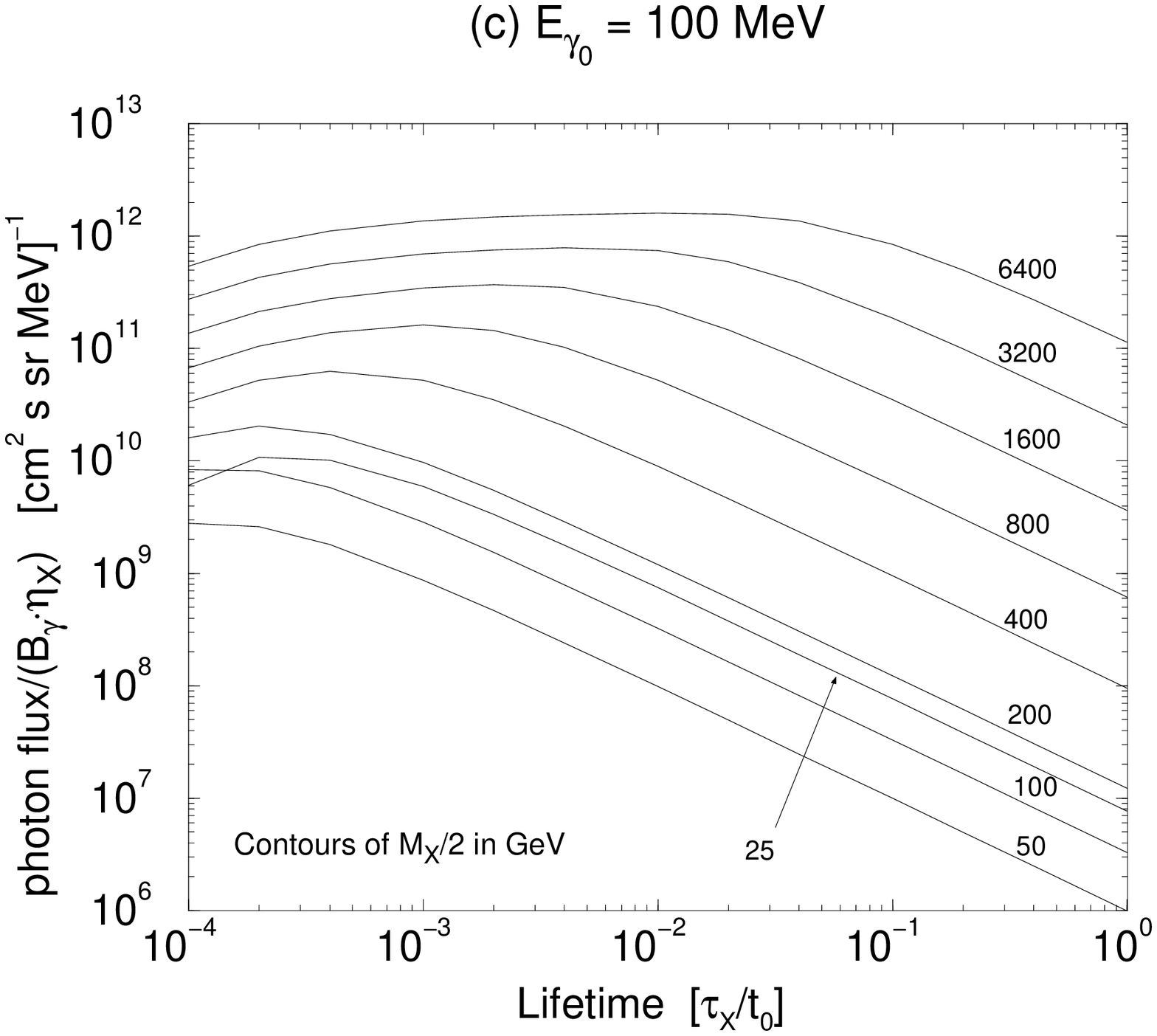}
\hfill
\epsfxsize=0.60\textwidth
\epsffile{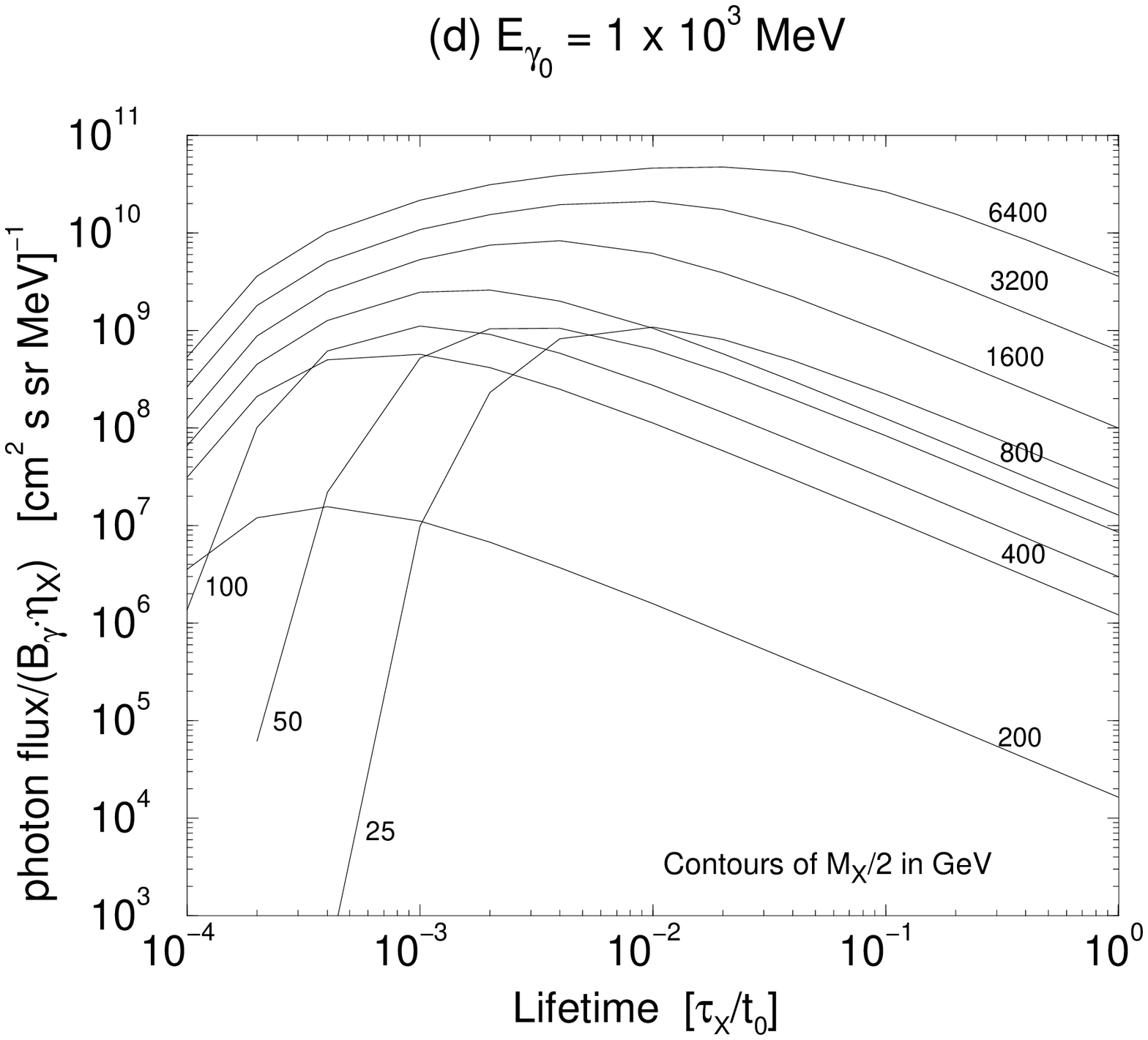}
\hfill }
\caption{Slices of Fig.~\ref{2-body-25-200-fig} with present-day 
detection energies $\Ephzero = 1$, $10$, $100$, $1 \times 10^{3}$~MeV\@.
To focus on the behavior of the photon flux for different masses
and present-day photon detection energies,
we restricted the lifetime $\tauX/t_0$ to be in the range $10^{-4} \ra 1$.
}
\label{2-body-lifetime-fig}
\end{figure}

\begin{figure}
\centerline{
\hfill
\epsfxsize=0.60\textwidth
\epsffile{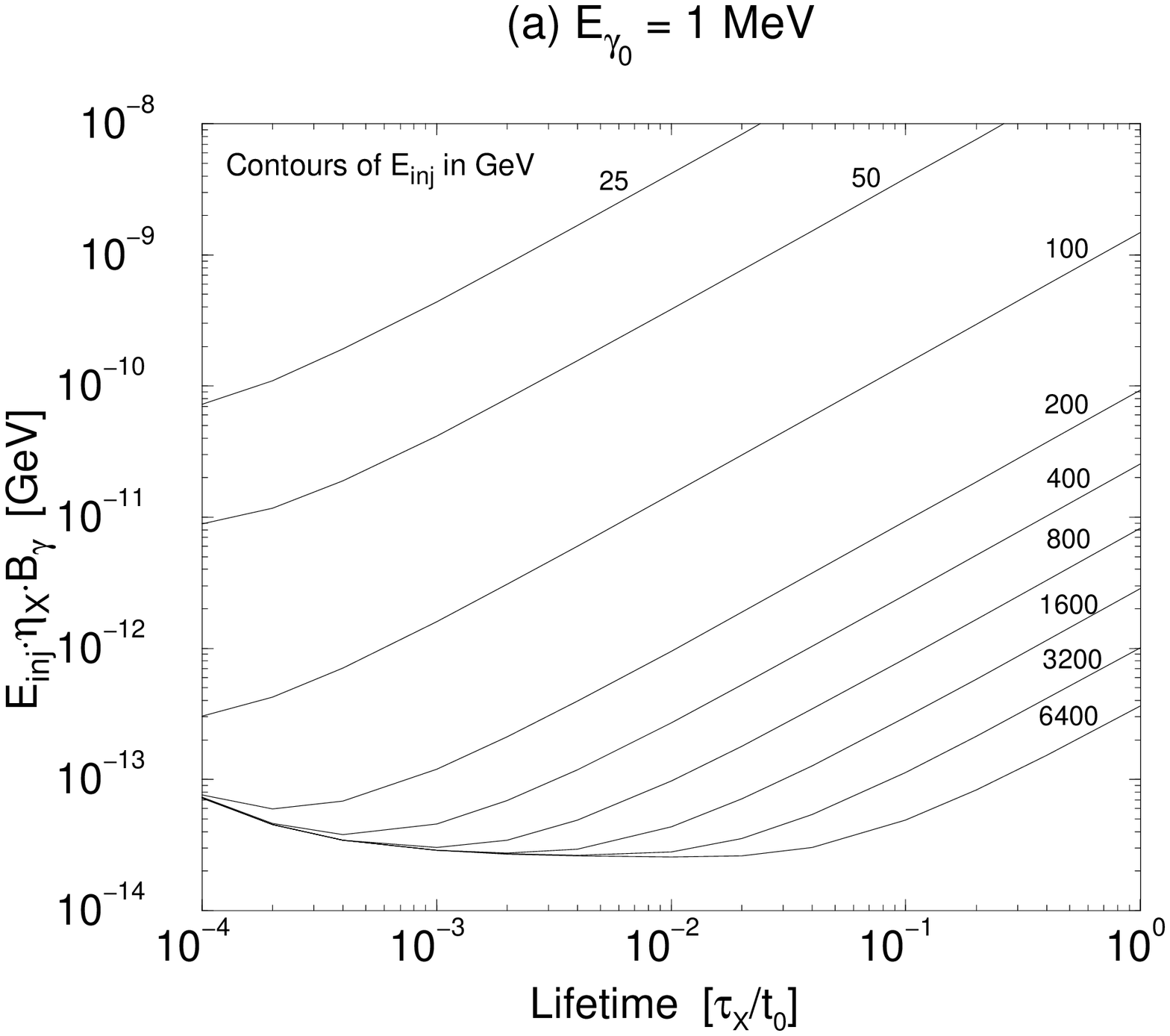}
\hfill
\epsfxsize=0.60\textwidth
\epsffile{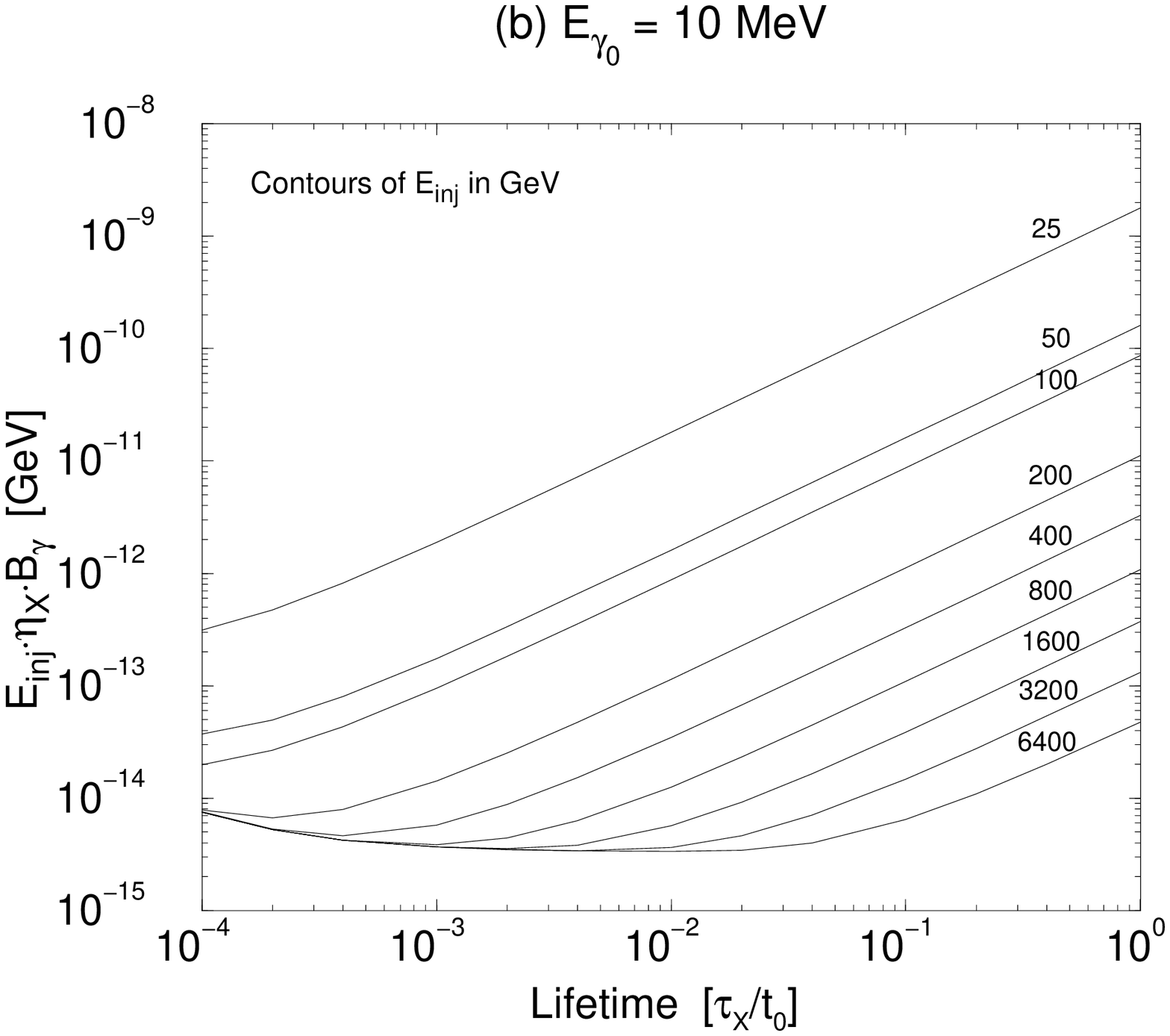}
\hfill }
\centerline{
\hfill
\epsfxsize=0.60\textwidth
\epsffile{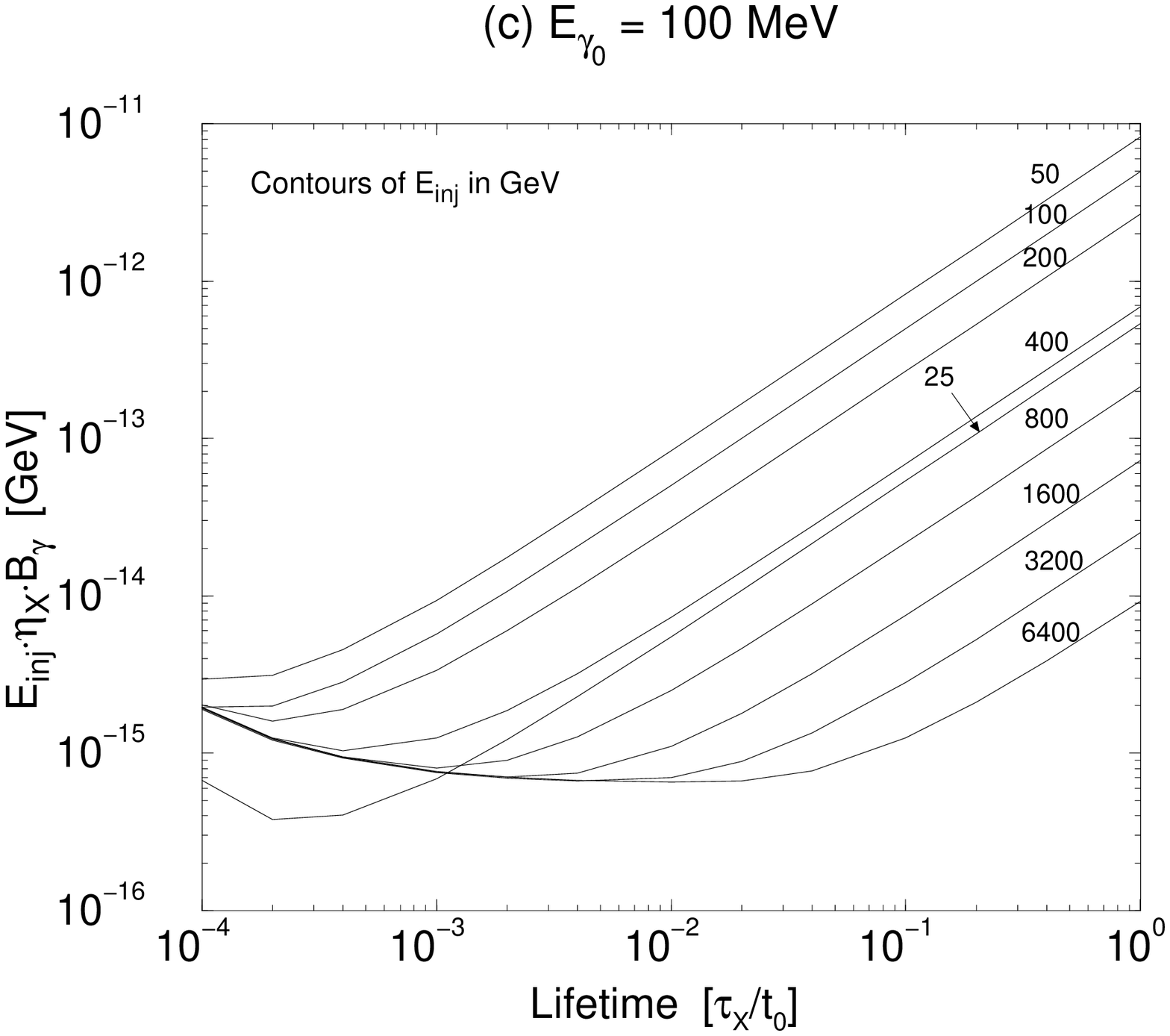}
\hfill
\epsfxsize=0.60\textwidth
\epsffile{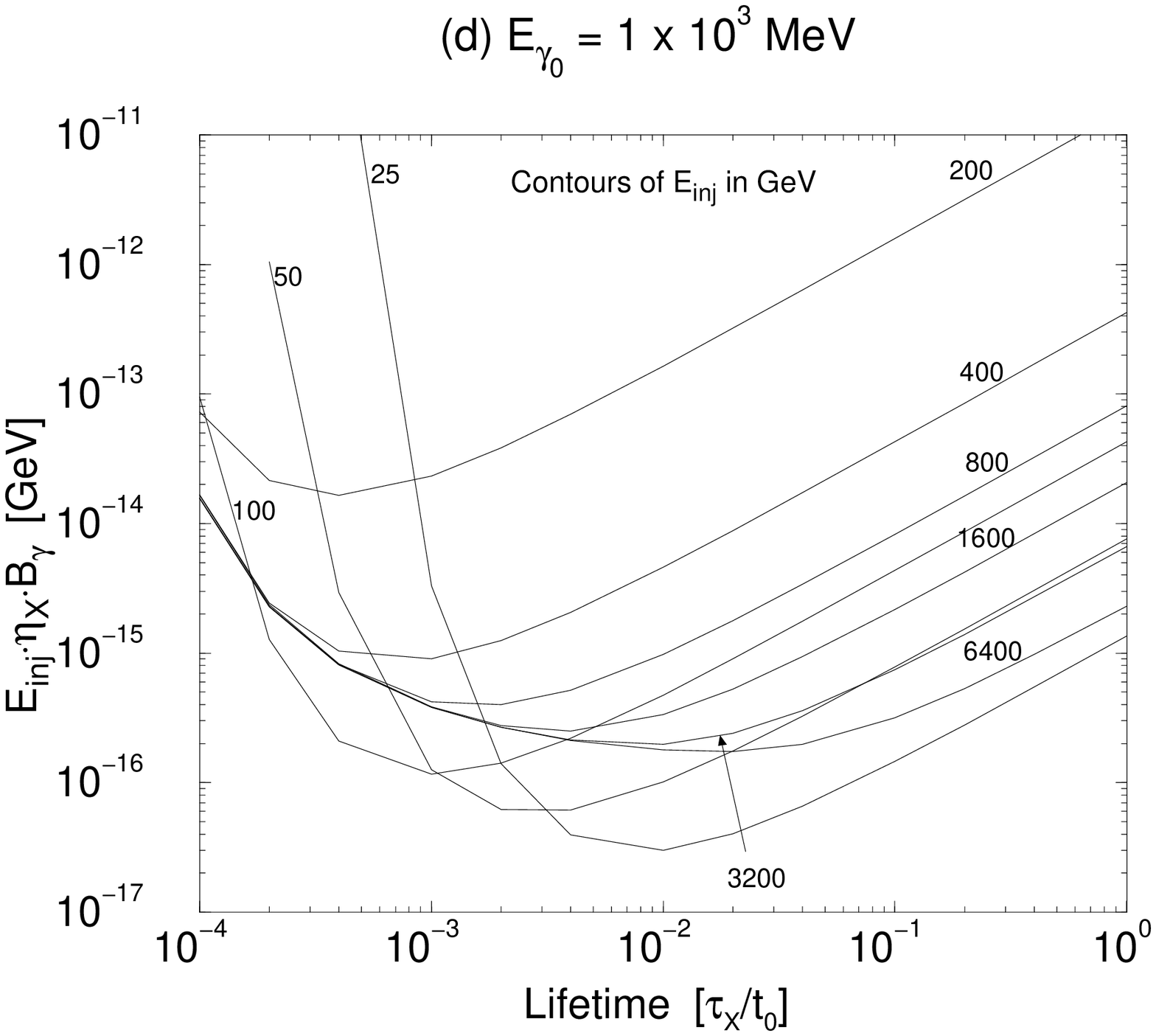}
\hfill }
\caption{Upper bounds on the relic density (times the radiative
branching ratio) for particular present-day detection energies
$\Ephzero = 1$, $10$, $100$, $1 \times 10^{3}$~MeV\@.
For a given relic mass, the region with relic density larger 
(or above) the mass contour is excluded.  Notice that better bounds
do not necessarily come from higher or lower detection energies.
}
\label{2-body-density-fig}
\end{figure}

In Fig.~\ref{2-body-lifetime-fig}, we have sliced the previous
photon flux vs.\ photon energy plots along the energy axis for a 
particular set of observed energies $\Ephzero = 1$, $10$, $100$, 
$1 \times 10^{3}$~MeV\@.  The bound on the relic density 
can be found by using the observational limit on the 
diffuse background found in Sec.~\ref{measurements-sec} 
for each photon energy $\Ephzero$, and effectively 
inverting the graphs in Fig.~\ref{2-body-lifetime-fig}
to give Fig.~\ref{2-body-density-fig}.  These figures
demonstrate that the best bound is not a trivial
function of the measured photon energy, relic mass or lifetime.
For example, in the $\Ephzero = 100$~MeV figure one 
finds a better bound on a $25$~GeV relic particle than 
for somewhat heavier relics. This is due to the fact that 
the unscattered photons will populate the higher energy range 
of the observed spectrum which is more strongly constrained. 
Whether or not such an inversion comes about depends upon 
whether the energy we are considering is larger or smaller than 
$\Edip$, defined in (\ref{edip-eq}).
For shorter lifetimes and larger photon energies,
one finds the unscattered part of the spectra is exponentially
suppressed, as can be seen in the lack of a limit 
for $\Ephzero = 1 \times 10^{3}$~MeV and small injected energies 
$\lsim 50$~GeV\@.

\begin{figure}
\centerline{
\epsfxsize=4.5in
\epsffile{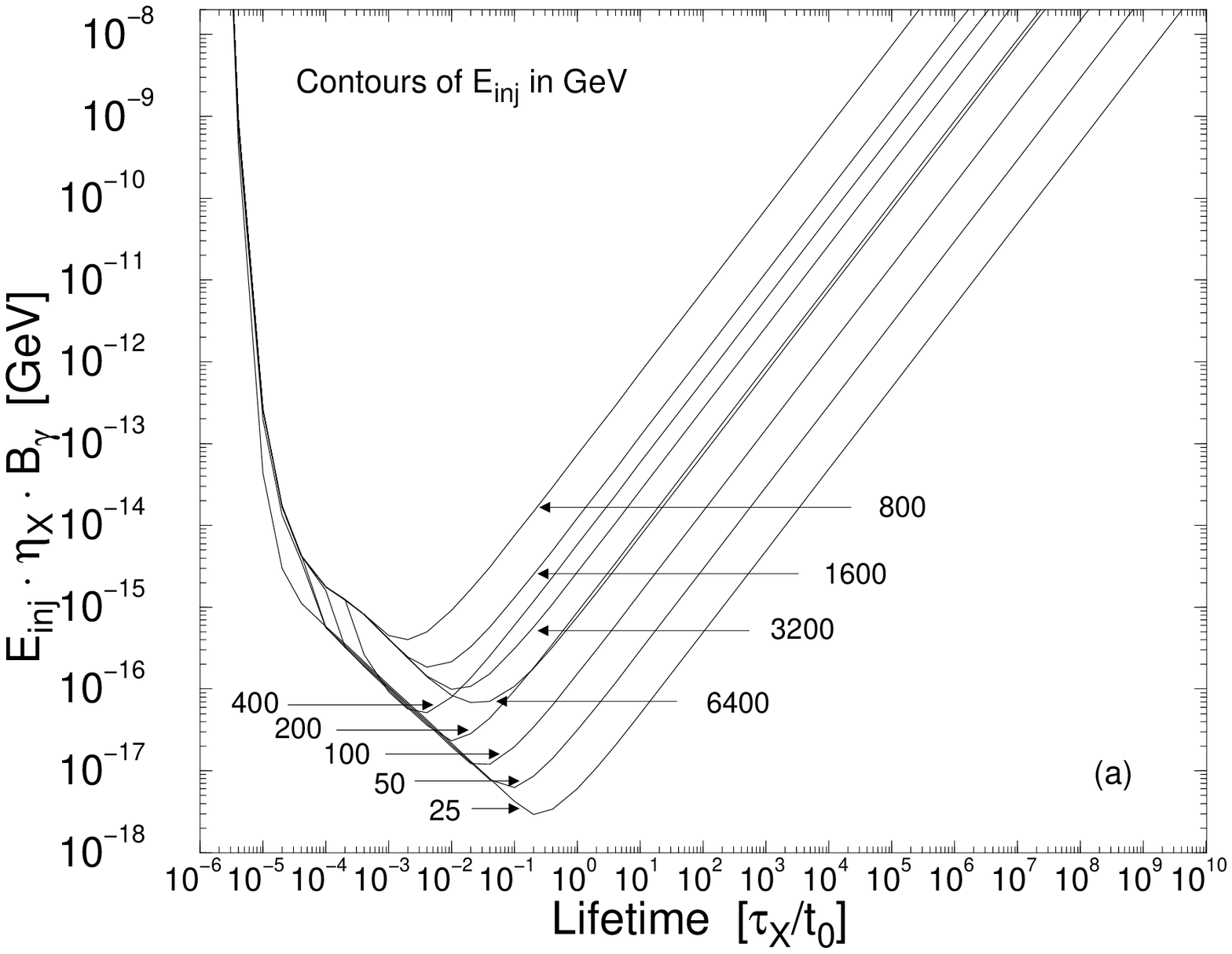}}
\centerline{
\epsfxsize=4.5in
\epsffile{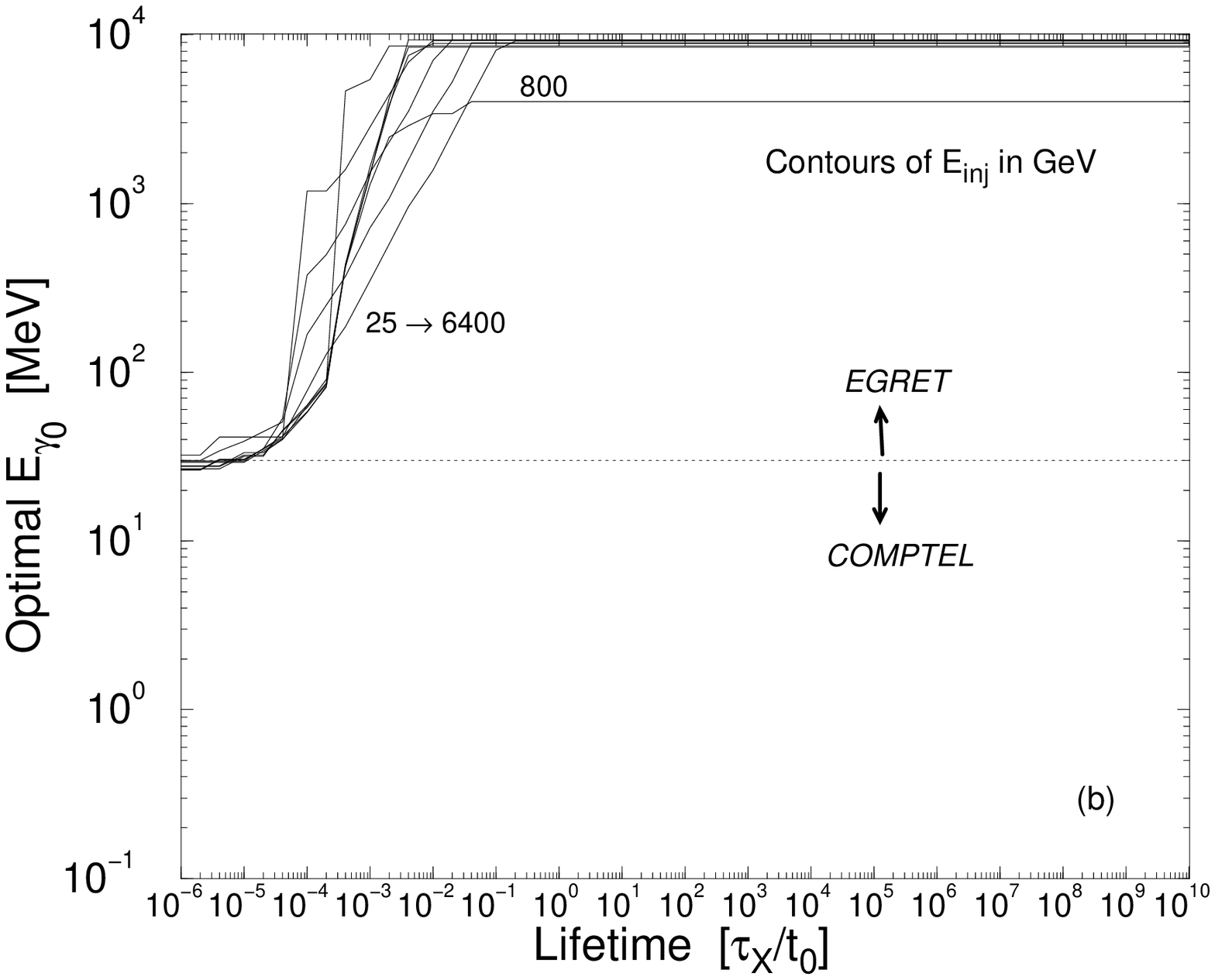}}
\caption{The upper graph shows the final relic density bound 
for 2-body radiative decays with lifetimes in the indicated range.  
The bound scales linearly with the radiative branching fraction 
of the relic $B_\ph$, although a branching ratio different from one 
does not strongly affect our bounds.  The upper limit on the relic 
density of $\sim 2 \times 10^{-8}$~GeV is roughly the critical density
corresponding to $\OmegaX h^2 \sim 1$.
The lower graph shows the optimal photon detection energy 
to obtain the best bound for a given lifetime.
This graph is divided at $\Ephzero = 30$~MeV with a dotted line
to show which instrument provides the diffuse photon background bound
for a given lifetime.
}
\label{2-body-bound-fig}
\end{figure}

Finding the maximum photon flux above background (i.e.\ one point
on each line in each graph of Fig.~\ref{2-body-25-200-fig}), allows one
to derive the bound on the relic density for a given mass and
lifetime as shown in the upper graph of Fig.~\ref{2-body-bound-fig}.
Bounds for lifetimes less than $\tauX/t_0 \approx 10^{-5}$ 
become poor very quickly due to the exponential suppression.  
Bounds for lifetimes longer than $\tauX/t_0 = 1$ scale by a 
factor $t_0/\tauX$ (outside (\ref{redshift-eq})) relative to the
bounds at $\tauX/t_0 = 1$.  We used the observational 
diffuse background fit as described 
in Sec.~\ref{measurements-sec}, and thus a bound for any given
mass and lifetime utilizes one (optimal) observational energy. 
This is shown in the lower graph of Fig.~\ref{2-body-bound-fig}.
For example, for $\tauX/t_0 \ge 1$ one can see the trend in 
increasing $\Ephzero$ is to increase the photon flux 
(see Fig.~\ref{2-body-25-200-fig}). Hence the best bound
for this lifetime comes from observations of the most energetic photons.
On the other hand, for $\tauX/t_0 = 10^{-4}$ one finds the
best bound for $\Einj = 25$~GeV is roughly $\Ephzero \approx 80$~MeV\@.
Higher $\Ephzero$ simply pushes into the exponential suppression
regime where no bound exists.  
The upper limit on $\MX\etaX \sim 2.5 \times 10^{-8}$~GeV corresponds 
to the critical density $\OmegaX h^2 \sim 1$, which is the
upper limit for any relic based upon the age of the universe. 

The general behavior in Fig.~\ref{2-body-bound-fig} is an 
increasing upper bound on $\MX\etaX$ as the injected energy is raised 
up to about $\Einj = 800$~GeV, and then a steady decrease
thereafter for larger injected energies.  The reason for
this trend in the bounds for 2-body decays is due to the 
transition noted in Table~\ref{2-body-crit-table} when the 
injected energy crosses $\Einj = 600$~GeV\@.  As remarked above, 
the value $\Edip = 10$~GeV at this transition
implies that for all injected energies above $600$~GeV, bounds can only 
be derived using the {\em scattered}\/ photons.  Since the number
of photons increases as the injected energy increases in the
scattering regime, the bounds are stronger as the energy is increased 
above $600$~GeV\@. This effect is further enhanced by
the fact that as the mass of the relic increases showering can occur
at smaller redshifts. As the injected energy
is reduced below $600$~GeV, more of the non-scattered, redshifted
photons appear, and so a better bound comes from lowering
the injected energy.  This is clear since the best bound
comes from the lowest injected energy considered, $\Einj = 25$~GeV\@.
The one special case is for $\Einj = 800$~GeV where the best 
bound for long lifetimes $\tauX/t_0 \gsim 1$ is not the 
highest present-day detection energy, but instead slightly
less $\Ephzero \approx 4 \times 10^{3}$~MeV\@.  The reason 
the bound comes from lower energies is due the $E_\ph^{-4.8}$
suppression in the scattered photons that exists for 
present-day energies within a factor of $\Emax/\Ecrit = 3$ 
lower than $\Edip = 17$~GeV\@.  Since the diffuse background
scales as $E_\ph^{-2.07 \; {\rm to} \; -2.38}$, the best
bound will be determined by the point where the redshifted spectrums' slope
is equal to the slope of the diffuse background . This point is roughly 
given by $\Edip/3$.

It is also interesting to note that the bounds for shorter 
lifetimes show scalings with the mass.  This can be seen 
by the fact that for shorter lifetimes the bounds begin to 
lay on top of each other in Fig.~\ref{2-body-bound-fig}. 
This scaling is seen to group into two lines: one line of 
which consists of the relics whose bounds come from 
scattering (i.e.\ $\Einj > 600$~GeV) and another line for
those relics whose photons can be directly detected.

\section{Bounds from Hadronic Decays}
\label{hadronic-decays-sec}
\indent

The bounds on radiative decays calculated above are strong, but 
it is not obvious that such decays ought to dominate the
branching ratio of heavy relics.  Here we establish bounds
on relic particles that decay through hadronic channels. 
We consider 3-body decays of relics into all kinematically 
available quark pairs and one uncolored (assumed massless) 
spectator.

\subsection{The photon spectrum from hadronic decays}
\label{photon-spectrum-subsec}
\indent

We assume general vector--axial couplings leading to both 
charged and neutral current mediated decays.  (CKM mixing 
is ignored since it is in general model dependent and can 
be absorbed into the couplings).
We take $m_t = 175~\GeV$, and so two thresholds exist with
increasing mass $\MX$ of the relic: $\MX = m_t + m_b$ 
(for charged current mediated decays) and $\MX = 2 \, m_t$
(for neutral current mediated decays).  Thus,
for a given mass $\MX$, an ensemble of relic decays 
with final quark energy and momenta spanning the 3-body phase space
can be constructed.

The quark pairs are fragmented and decayed according to the string 
fragmentation scheme~\cite{string-fragmentation} implemented 
in {\sc Jetset}~\cite{Jetset}.  
This fragmentation scheme has been well tested with collider 
experiment data, and the resulting photon spectrum from 
{\sc Jetset} is discussed in Ref.~\cite{BengGamma}.
In particular, the exact of shape and normalization of the 
photon spectrum depends on the particular final state 
quarks~\cite{BengGamma} but generally scales with
the relic mass $\MX$, which we discuss below.  
Once the spectrum is calculated, the present-day photon flux
can be determined from (\ref{redshift-eq}).

The 3-body decay allows energy to escape with the
uncolored decay product. Hence, the integrated photon spectrum
appearing after the hadronization and decay of 
the $\qqbp$ system is always less than $\MX$.  Furthermore,
the hadronization process does not uniquely end with 
neutral pions, as some energy leaks into leptons.  
We consider only the effects of pion decays into photons.
Therefore, the fraction of energy appearing in the final state
(before redshifting) is generally between about $\MX/10$
and $\MX/3$ on average.  This is to be contrasted with
the 2-body radiative decays, where the total photonic energy
injected is equal to the mass of the relic.

\subsection{Generating the photon spectrum}
\label{generating-subsec}
\indent

The photon spectrum is obtained directly from the hadronization 
of different final state quark pairs and is presented in 
Fig.~\ref{compare-fig}.  As we discuss below, charged
current and neutral current mediated decays yield nearly
identical results (except near the thresholds associated with
producing one or two top quarks). As such, we will only consider
neutral current decays.  In total, $10^4$ events were generated 
for each final state quark pair, with
each event corresponding to a point in the 3-body decay 
phase space\footnote{Since the energy of the $q\overline{q}$
system has the typical 3-body distribution, a direct comparison
cannot be done between our results and those of an annihilation signal 
with fixed energy as done in Ref.~\cite{BengGamma}.
However, we have checked that in the appropriate limit
we recover the shape and normalization found there.}.
We have included final state electromagnetic and QCD radiation 
showers prior to fragmentation, although the showering is
performed only off the final state quark pair.  
The particular quark pair $t\overline{t}$ is a special
case since the top quark decays before it fragments.
However, the photon spectrum is not particularly sensitive
to Monte Carlo ordering of decay vs.\ fragmentation
({\sc Jetset} fragments before decay).
In particular, we compared the photon spectrum produced 
from the {\sc Pythia} process $e^+e^- \rightarrow t\overline{t}$
after fragmentation (with no initial state radiation) 
at large energies $\sqrt{s} > 2 \, m_t$
with the top quark decaying before and after fragmentation.
The photon spectrum is slightly enhanced for energies
$E_\ph / \sqrt{s} > 0.05$ when the top quark is allowed
to decay before it fragments.  However, the dominant 
effect on our bounds from relic decays into heavy top quarks 
comes from the large number of photons produced at {\em low}\/ 
energies.  In fact, Fig.~\ref{compare-fig} clearly 
shows a much larger number of photons from the hadronization of 
light quarks ($d\overline{d}$, $u\overline{u}$, etc.) 
at energies $E_\ph/\MX \gsim 0.01$ 
where one would roughly expect the slight enhancement 
in the top quark photon spectrum.  Thus in regimes where the 
higher energy photons define the bound, it is the light quarks' 
photon spectrum that is crucial.
 
The photon spectrum originates almost entirely from decaying pions, 
which are created in the quark fragmentation 
process~\cite{string-fragmentation,BengGamma}.  Thus, for a given decay 
$\MX \ra q\overline{q}$, the spectra 
scale with $\MX$ and can be normalized to the mass of the 
relic.  It is only the finite mass of the 
pion that breaks the scaling behavior.  The photon spectrum 
is also virtually independent of the vector--axial 
couplings in the 3-body phase space.  The only energy dependence
enters in the mass of the uncolored product which  we take 
to be negligible compared with the mass of the relic.
In fact, a large non-zero uncolored product mass could
be easily accommodated by simply reducing the relic mass by 
approximately the mass of the uncolored product.

Once the photon spectrum has been obtained from {\sc Jetset}, 
it is fit to a sum of exponentials~\cite{BengGamma}
\begin{equation}
\LL( E_\ph ) \times \MX \; = \; \frac{dN_\ph}{dx} \; = \; 
    A \, e^{-\alpha x} \, + \> B \, e^{-\beta x},
\label{exponential-fit-eq}
\end{equation}
where $x \equiv E_\ph/\MX$, and $A$, $B$, $\alpha$, $\beta$
are positive constants.  The fit gives a reasonable 
characterization of the photon spectrum valid for $E_\ph \gsim m_{\pi}/2$.  
The dependence of the fit on the Monte Carlo statistics is small.
Increasing the statistics by a factor of 5 shifts
the final redshifted spectra by at most ${\cal O}( 15\% )$.

\subsection{Numerical results: Photon flux versus photon energy}
\indent

\begin{figure}
\centerline{
\hfill
\epsfxsize=0.60\textwidth
\epsffile{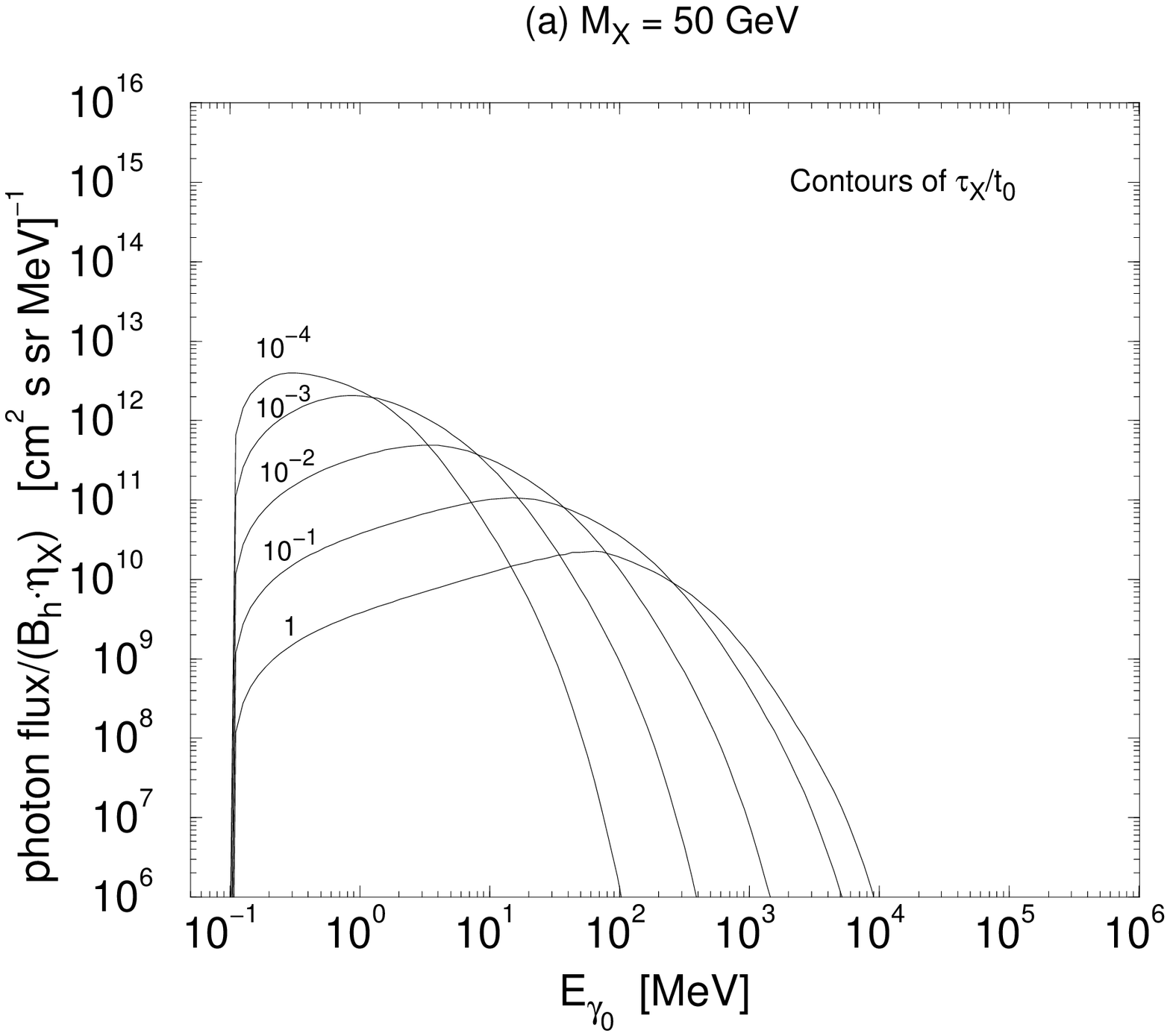}
\hfill
\epsfxsize=0.60\textwidth
\epsffile{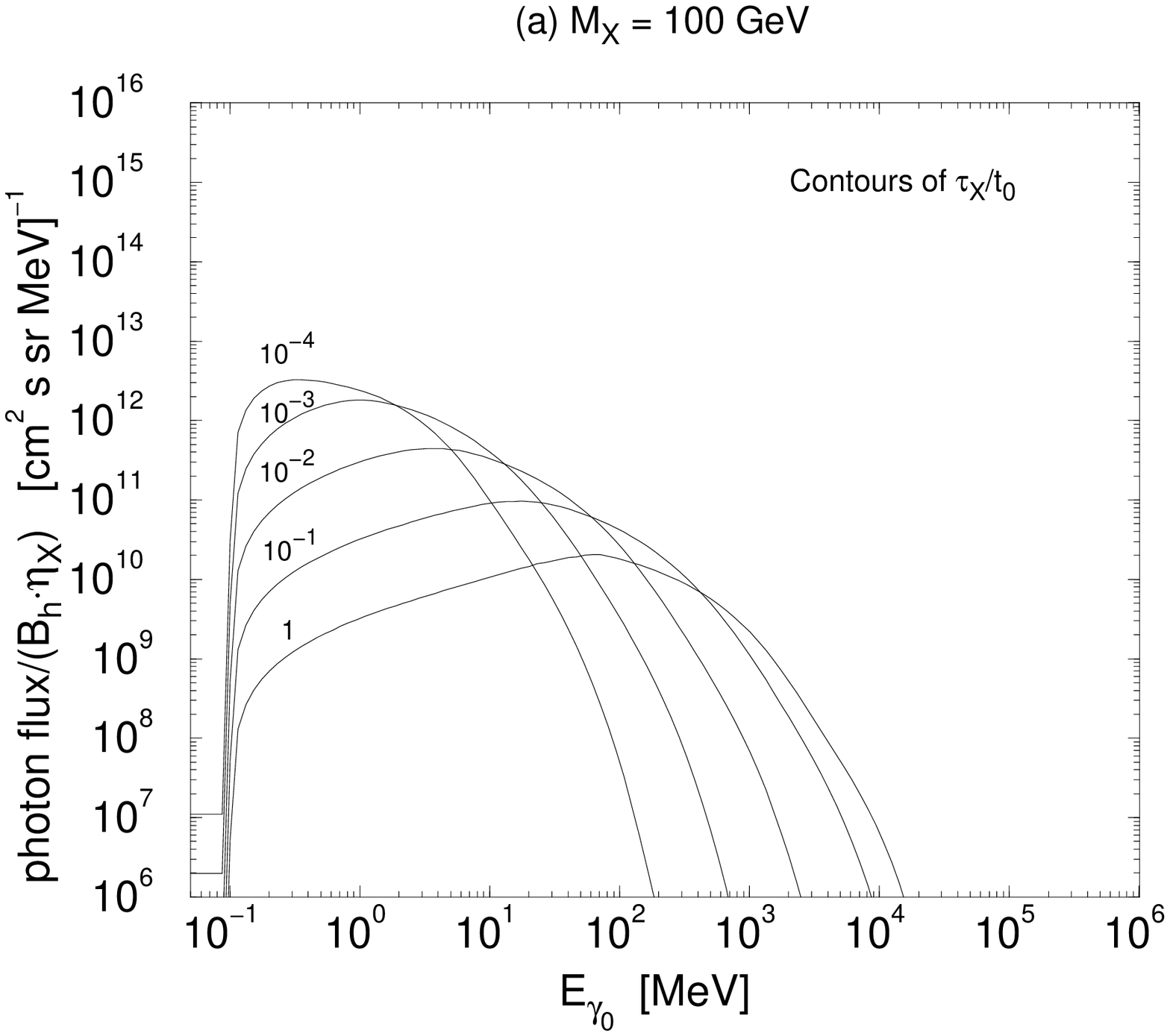}
\hfill }
\centerline{
\hfill
\epsfxsize=0.60\textwidth
\epsffile{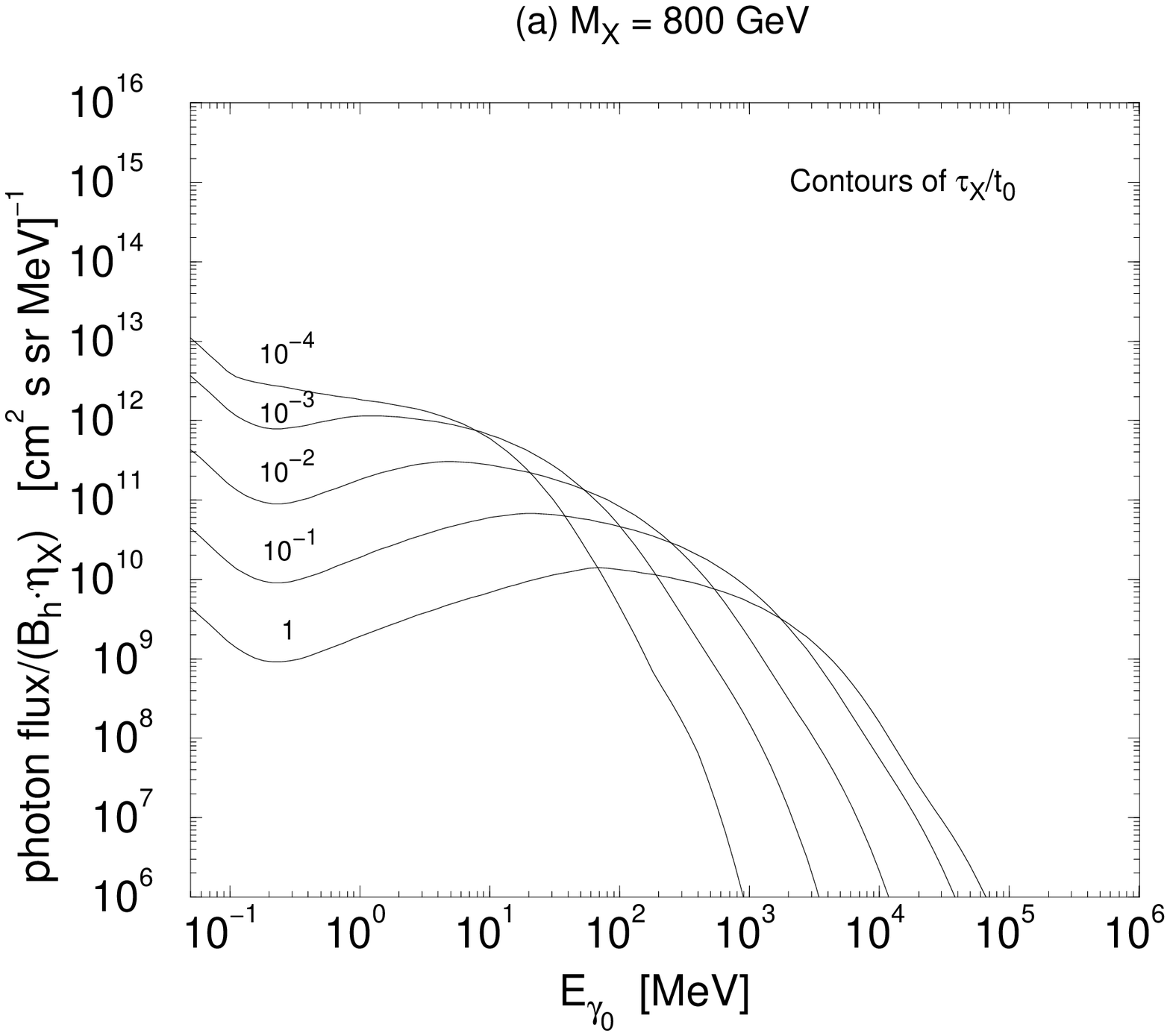}
\hfill
\epsfxsize=0.60\textwidth
\epsffile{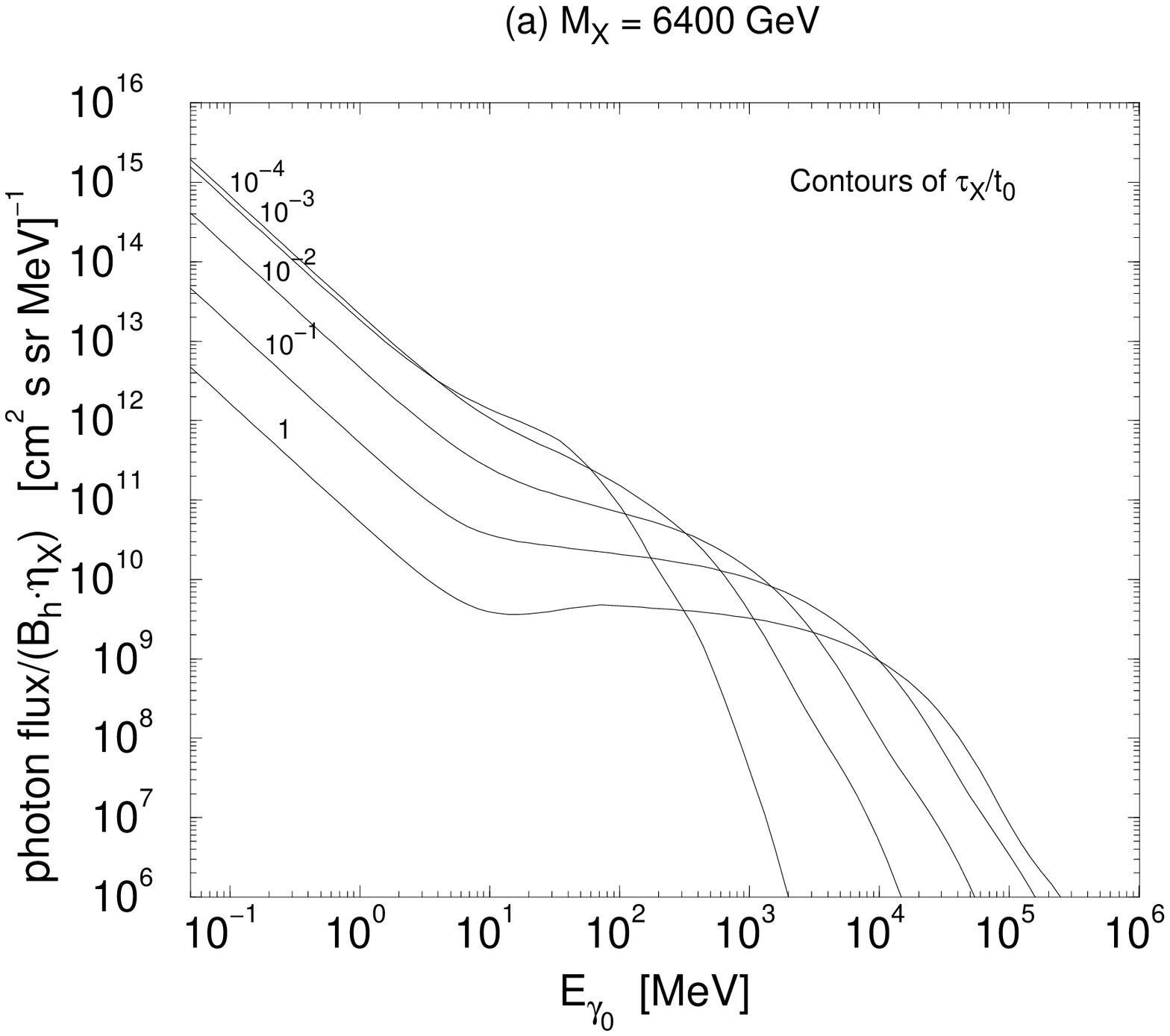}
\hfill }
\caption{Redshifted photon spectra for a relic with mass
$\protect\MX = 50$, $100$, $800$, $6400$~GeV, 
a hadronic branching ratio $B_h$, and a relic density $\etaX$.
All kinematically allowed 3-body hadronic 
decays $X \ra q\overline{q}$ ($+$ uncolored product) 
are assumed to occur with equal branching ratio.
A sample of relic lifetimes $\tauX/t_0$ were chosen 
and plotted as separate contours on the graph.
}
\label{hadronic-50-400-fig}
\end{figure}

We have scanned the relic mass range over more than two orders of 
magnitude from $\MX = 50$~GeV (a likely lower bound from LEP) 
through $12.8$~TeV in steps of a factor of $2$.  We have 
simultaneously scanned the lifetime range throughout the
region that gives bounds for relevant densities. 
  Fig.~\ref{hadronic-50-400-fig} displays a selection of
the above sampling, with relic masses $\MX = 50$, $100$, 
$800$, $6400$~GeV and lifetimes including $\tauX/t_0 = 10^{-4}$, 
$10^{-3}$, $10^{-2}$, $10^{-1}$, $1$.  The figures
show the photon flux as a function of the present-day
photon energy.
We have also calculated the photon flux as a function of the 
photon energy for the same relic mass--lifetime parameter space 
using charged current interactions.  The photon fluxes are virtually
identical throughout most of the mass range, with the neutral current 
interaction usually
giving a slightly larger value (due to decays into top quark pairs) 
than decays via charged 
current interactions.
However, in the mass window $m_t + m_b < \MX < 2 \, m_t$ 
the photon flux from a charged current interaction is 
about a factor of two greater than the equivalent spectra from
a neutral current interaction, since the 
$b\overline{t}$ decay mode is open.

It is clear from Fig.~\ref{compare-fig} that most of the 
photons are well below $\MX/10$, so that the effect of 
scattering for hadronic decays is not as prominent as in radiative 
decays for the same relic mass.  The effects of photon--photon
scattering and pair production are handled similarly to
the radiative decays, however the input spectra
is no longer a delta function in the photon energy.  Specifically, 
the spectra for photon--photon scattering~(\ref{GG-eq}), 
pair production at low $z$~(\ref{PP-lowz-eq}) and pair production 
at high $z$~(\ref{PP-highz-eq}) are used in the same way
as in radiative decays.   We need only
determine how much energy is injected into
each regime.  Since the energy injected into this regime is dependent
on $\Emax$ and $\Ecrit$, which are in turn dependent on
the redshift $z$, the procedure must be done numerically.

As noted above the non-scattered spectra are cutoff at
$E_\ph = m_\pi/2$, whereas the scattered spectra
have no such cutoff.  Thus, in regimes where the non-scattered
(injected) spectra dominate the final redshifted spectra, 
we expect a cutoff 
at $\Ephzero = \frac{m_\pi}{2}/(1+z_0) \approx 10^{-1}$~MeV\@.
Such a cutoff is observed in all of the spectra
in Fig.~\ref{hadronic-50-400-fig}. 
Fig.~\ref{hadronic-50-400-fig} also shows explicit scaling 
with relic mass in the redshifted spectra.
This is a consequence of the string fragmentation process, 
which is Lorentz invariant (given that enough energy is
present to create a string and subsequently, jets).
Scaling does not hold, however, for the scattered spectra
since there are absolute cutoffs ($\Emax$, $\Ecrit$) involved.
Note also that despite the fact that there is no scattering, 
the location of the peak photon flux increases above $10^{-1}$~MeV 
as the lifetime is increased.  This is to be expected since as the 
lifetime $\tauX/t_0 \ra 1$, the peak flux should approach $m_{\pi}/2$ 
since photons that have not been appreciably redshifted ($z \sim 1$)
are not exponentially suppressed.

Different final state quark pairs give rise to different 
final spectra.  However, the
principle differences between decays into particular 
final state quark pairs is not difficult to understand.
As can be anticipated from the injected (non-redshifted)
spectra in Fig.~\ref{compare-fig}, decays into top
quark pairs yield the best bound when the lowest
energy photons from the injected spectra are sampled.
The decay into top quark pairs gives the largest flux
of photons for relics with a large mass $\MX > 2 m_t$
at very low present-day energies $\Ephzero \ll \MX/(1+z_0)$.  
Similarly, it is the decay into light quark pairs $d\overline{d}$ and 
$u\overline{u}$ that give the largest flux of photons for
higher energy photons $\Ephzero \gsim \MX/(1+z_0)$.  Without 
a theoretical motivation for decays into one or another flavor 
or family, we choose to divide the branching ratio equally among 
the kinematically available quark pairs.  Thus we assign an 
equal branching ratio for all the pairs ($1/5$ or $1/6$ 
depending on whether the $t\overline{t}$ threshold 
has been crossed).

The effect of scattering on the spectra, as remarked near
the beginning of this section, is not as important for hadronic 
decays as it is for radiative decays.  In fact,
scattering is virtually absent for $\MX = 50$~GeV, as 
illustrated in Fig.~\ref{hadronic-50-400-fig} where there 
is no photon flux (above the lower limit in the graph)
for $\Ephzero < 10^{-1}$~MeV\@.  This is not surprising
since the quark pair will always have an invariant mass
less than $\MX/2$, which is only barely above the threshold 
for photon--photon scattering $\simeq 23$~GeV\@.
For $\MX = 100$~GeV, only
photon--photon scattering is possible, and one can
see the characteristic limiting behavior of a flat spectra 
for $\Ephzero < 10^{-1}$~MeV\@.  For $\MX \ge 200$~GeV, 
pair production dominates the scattered piece of the
redshifted spectra $\Ephzero < 10^{-1}$~MeV, albeit
with a total integrated energy that is much less
than the unscattered piece ($\Ephzero > 10^{-1}$~MeV).
It is really only when $\MX \gsim 1$~TeV that
the scattered piece of the spectra has a photon 
flux comparable to the unscattered piece.  This implies
that the bulk of the injected photons are below 
about $\MX/10$, which is roughly the scale where
scattering turns on.

\subsection{Relic density bounds}
\indent

\begin{figure}
\centerline{
\hfill
\epsfxsize=0.60\textwidth
\epsffile{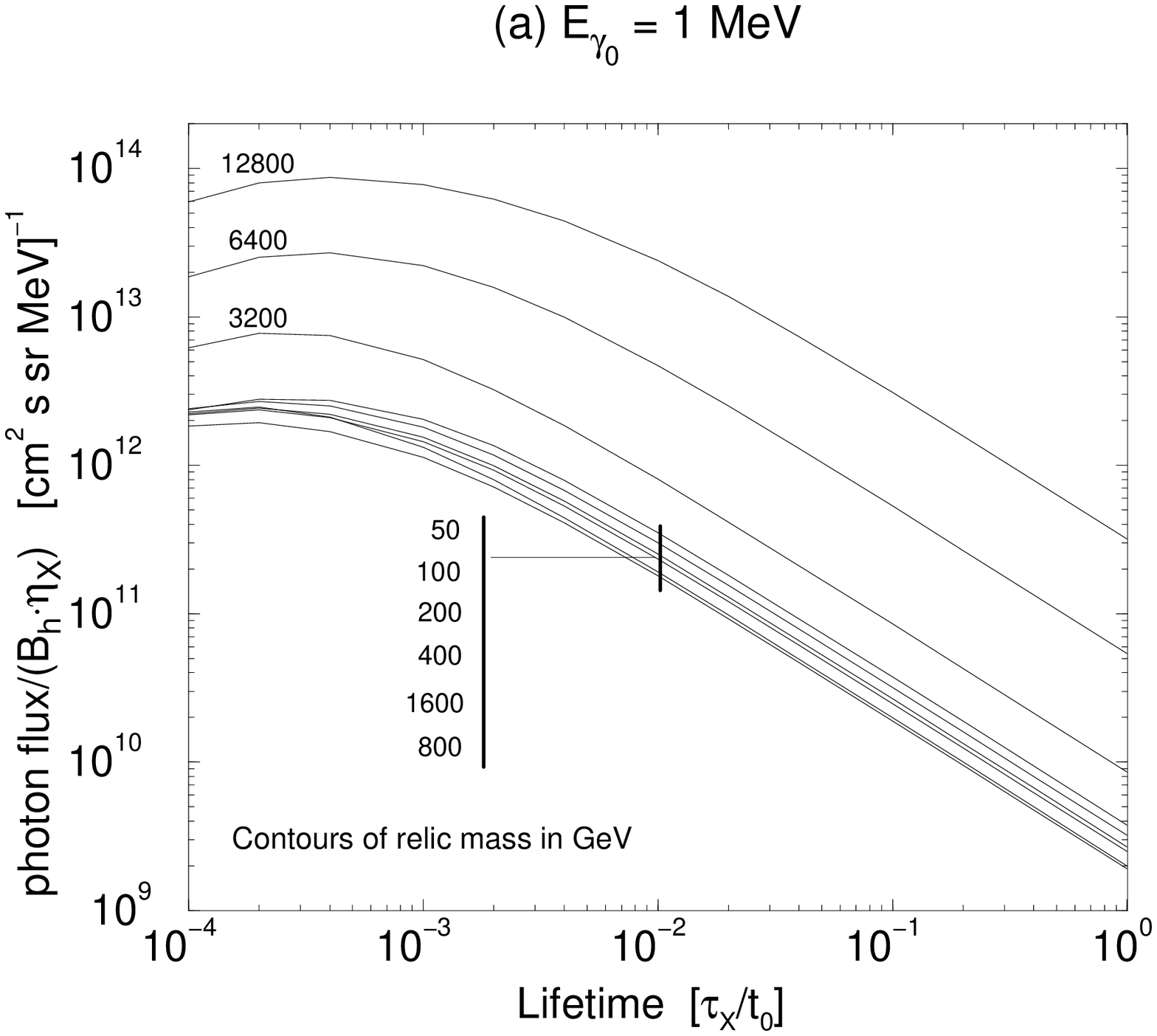}
\hfill
\epsfxsize=0.60\textwidth
\epsffile{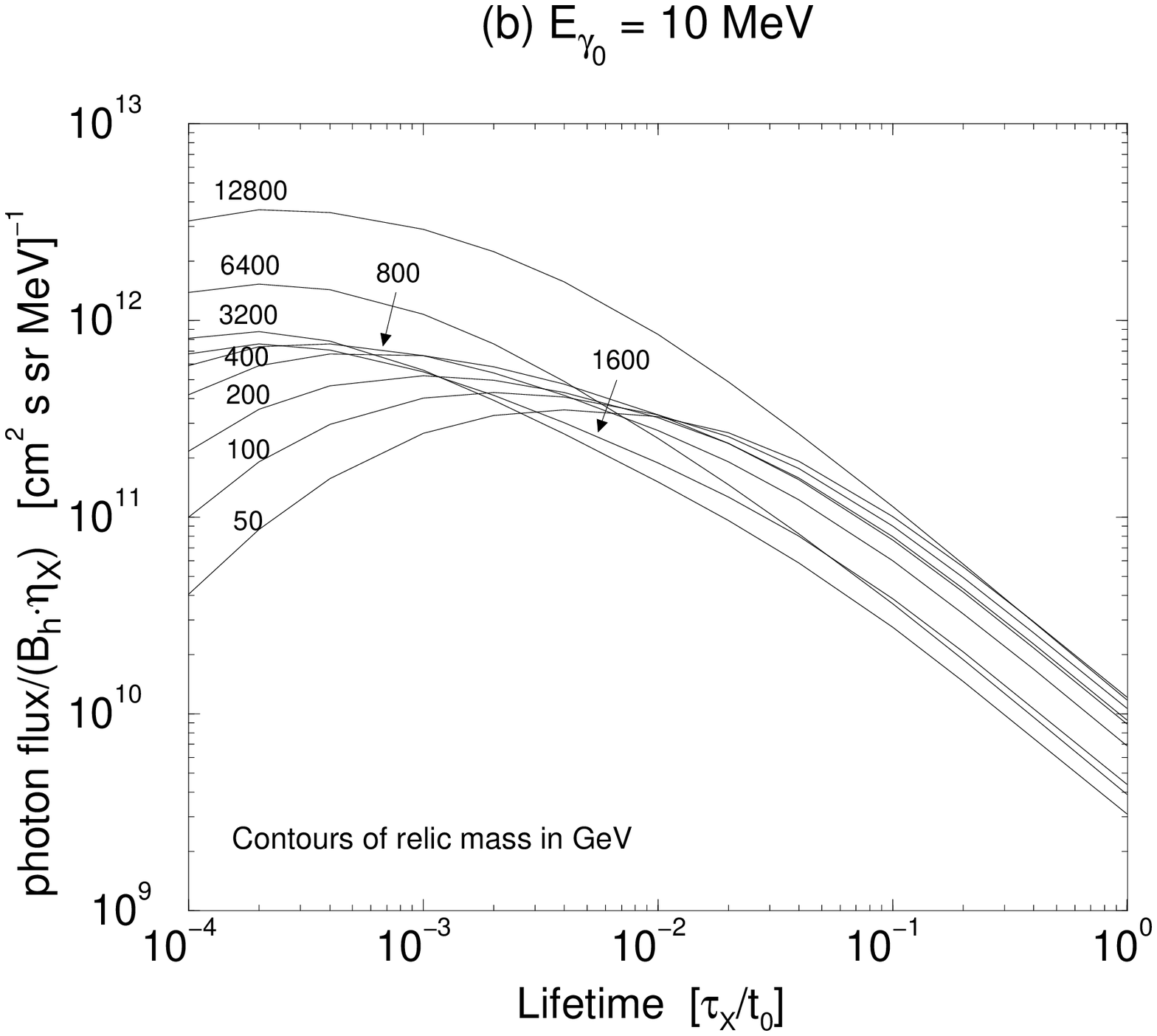}
\hfill }
\centerline{
\hfill
\epsfxsize=0.60\textwidth
\epsffile{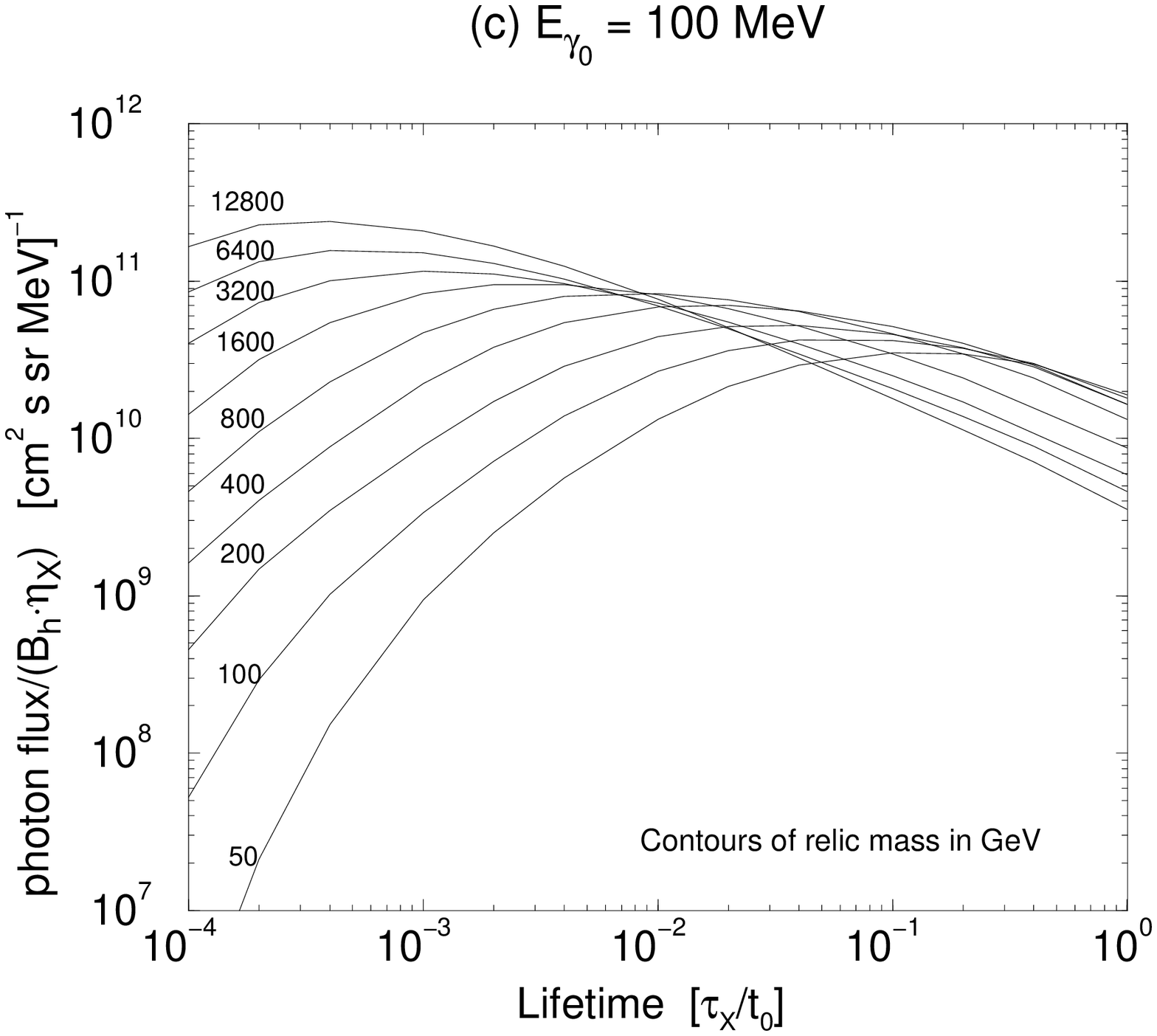}
\hfill
\epsfxsize=0.60\textwidth
\epsffile{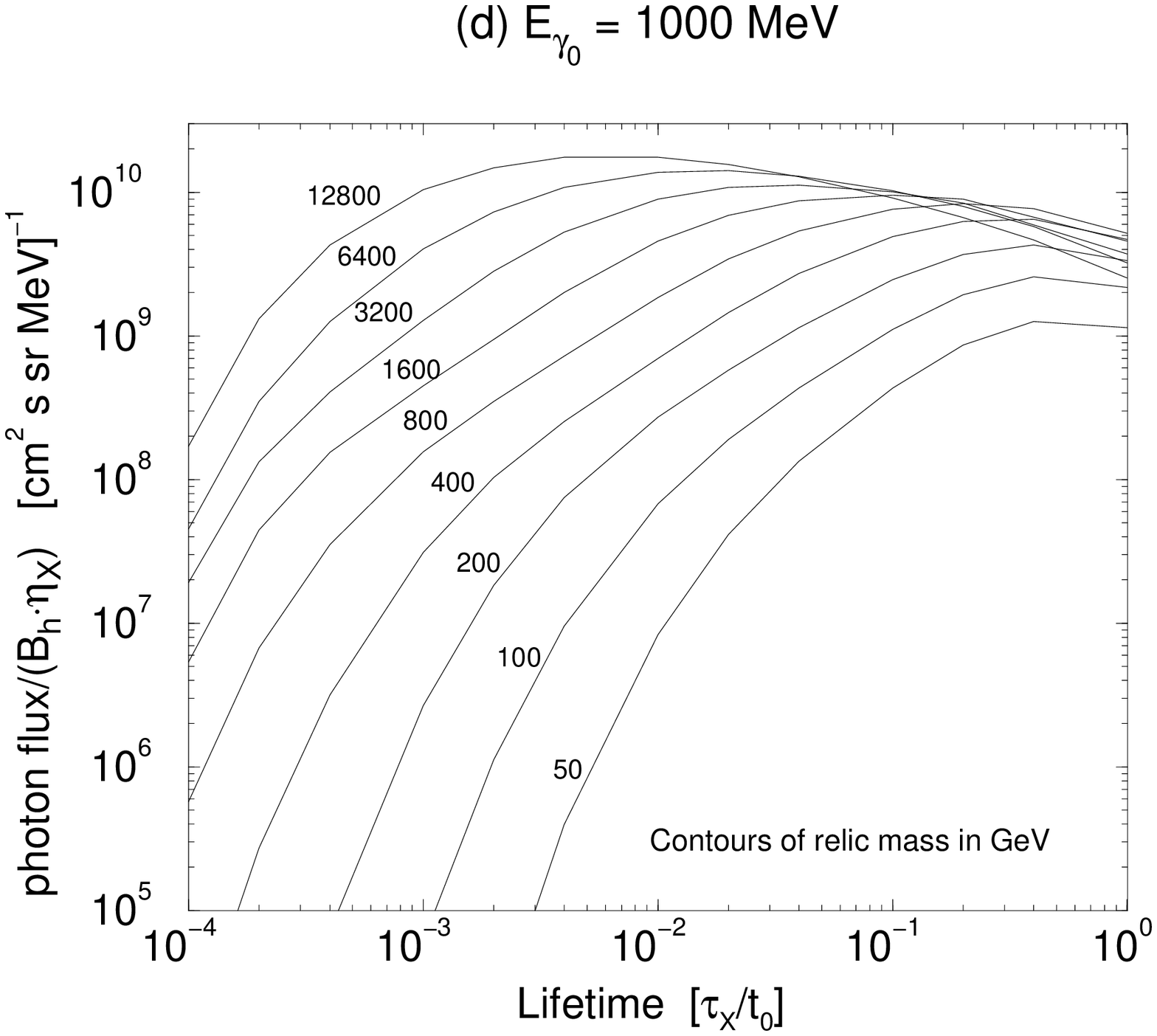}
\hfill }
\caption{Slices of Fig.~\ref{hadronic-50-400-fig} with present-day 
detection energies $\Ephzero = 1$, $10$, $100$, $1 \times 10^{3}$~MeV\@.
To focus on the behavior of the photon flux for different masses
and present-day photon detection energies,
we restricted the lifetime $\tauX/t_0$ to be in the range $10^{-4} \ra 1$.
}
\label{hadronic-lifetime-fig}
\end{figure}

\begin{figure}
\centerline{
\hfill
\epsfxsize=0.60\textwidth
\epsffile{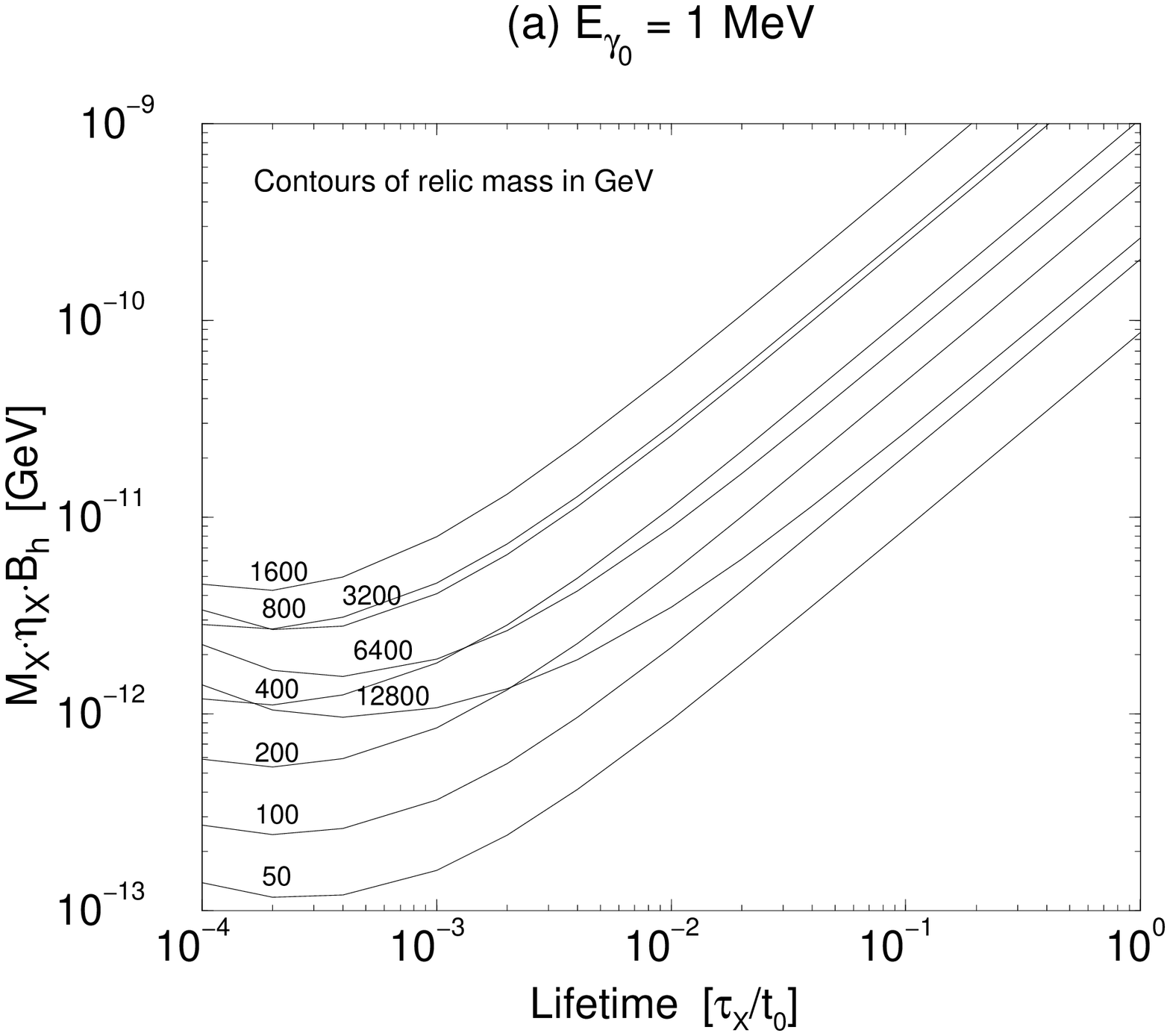}
\hfill
\epsfxsize=0.60\textwidth
\epsffile{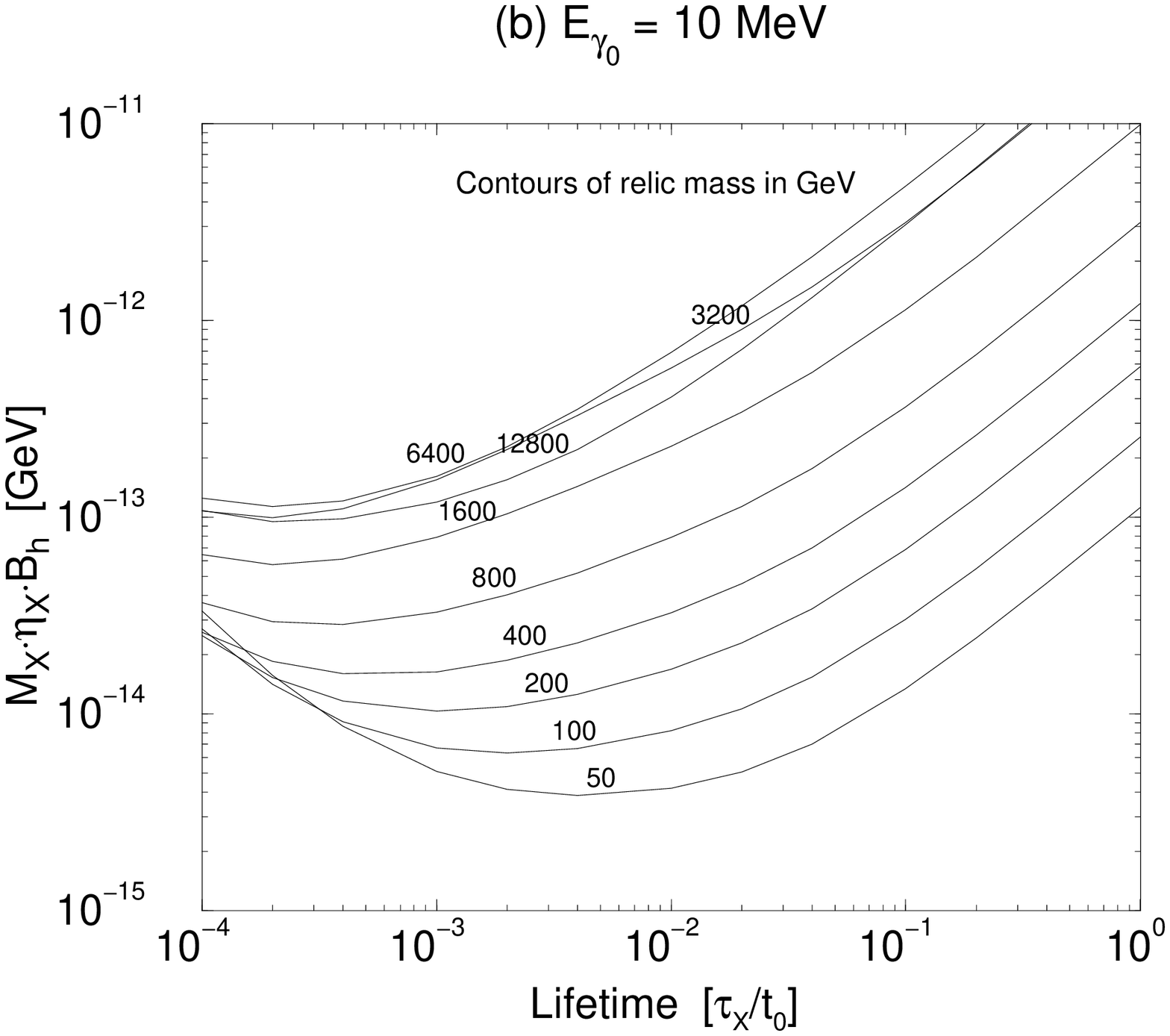}
\hfill }
\centerline{
\hfill
\epsfxsize=0.60\textwidth
\epsffile{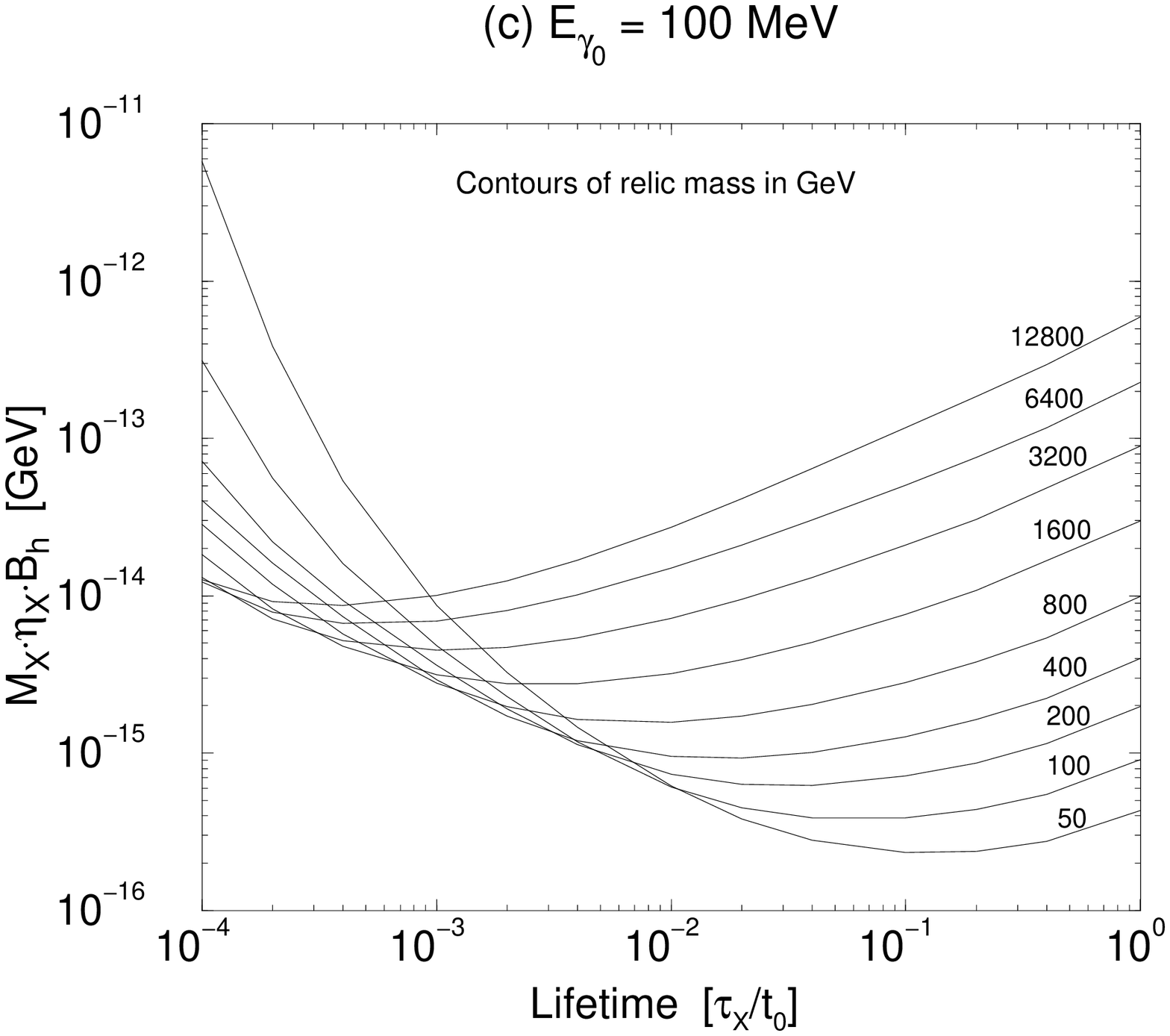}
\hfill
\epsfxsize=0.60\textwidth
\epsffile{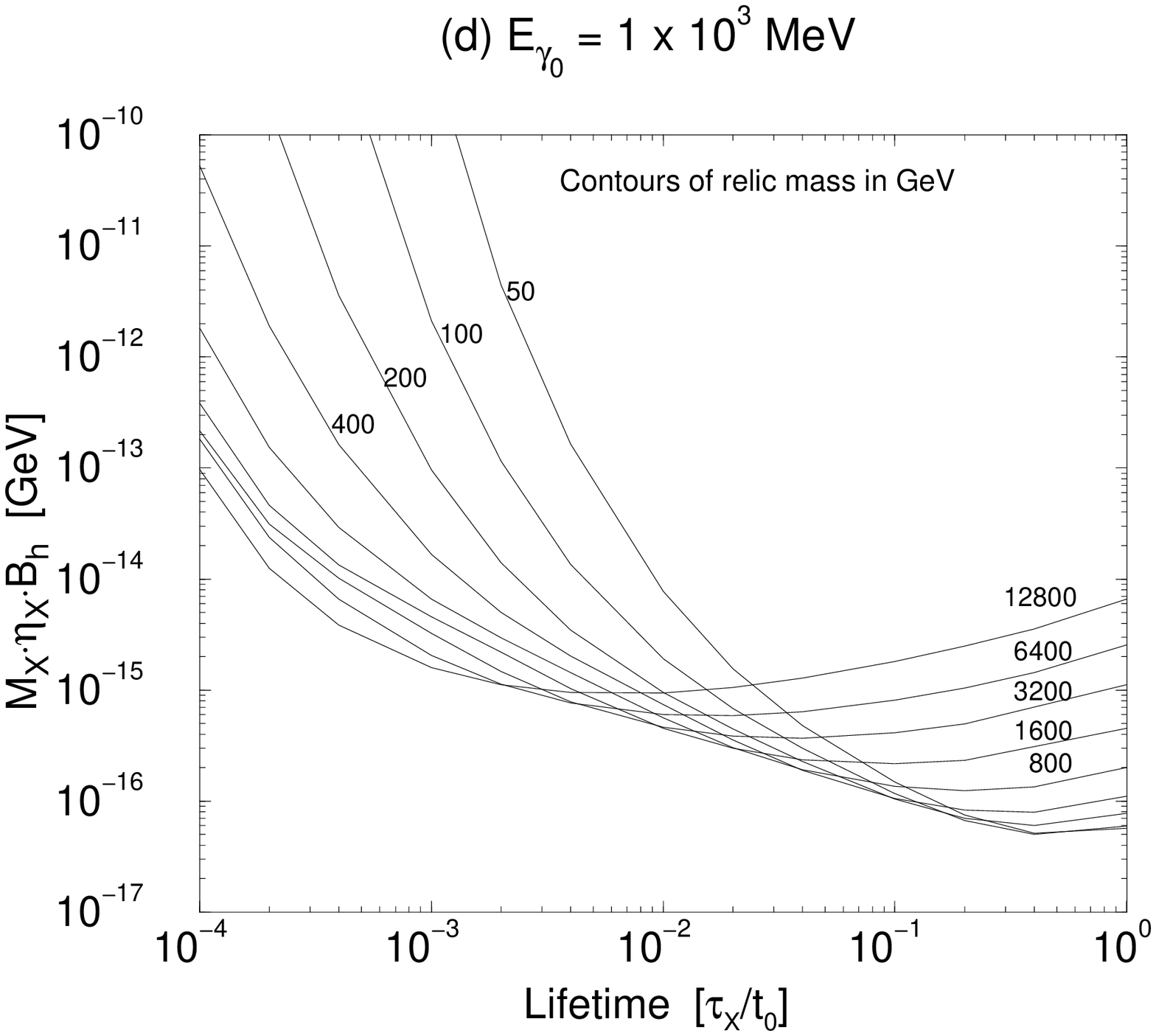}
\hfill }
\caption{Upper bounds on the relic density (times the hadronic
branching ratio) for particular present-day detection energies
$\Ephzero = 1$, $10$, $100$, $1 \times 10^{3}$~MeV\@.
For a given relic mass, the region with relic density larger 
(or above) the mass contour is excluded.  Notice that better bounds
do not necessarily come from higher or lower detection energies.
}
\label{hadronic-density-fig}
\end{figure}

In Fig.~\ref{hadronic-lifetime-fig} we have sliced
the previous photon flux vs.\ photon energy plots along
the energy axis in analogy to Fig.~\ref{2-body-lifetime-fig},
for the particular present-day energies $\Ephzero = 1$, $10$, $100$, 
$1 \times 10^{3}$~MeV\@.  Just as in the 2-body case, the bound on the 
relic density can be found by using the observational limit 
on the diffuse background found in Sec.~\ref{measurements-sec} 
for each photon energy $\Ephzero$, as shown in
Fig.~\ref{hadronic-density-fig}.  We observe, as in the 2-body 
radiative decays, that the mass dependence shows nontrivial behavior 
characteristic of the transition between 
relics whose bounds come the scattered and unscattered 
spectra respectively, for $\MX \gsim 1$~TeV\@.
In addition, one finds that for shorter lifetimes and larger photon 
energies the spectra are exponentially suppressed, as can be seen 
in the lack of a bound for $\Ephzero = 10^{3}$~MeV and small 
masses $\MX \lsim 100$~GeV\@.   The physical interpretation
is that most of the decays occurred much earlier than our 
present epoch,  so the photon flux is significantly more 
redshifted than for relics with a longer lifetime.  Thus, we 
see a much smaller number of present-day photons at high energy.

\begin{figure}
\centerline{
\epsfxsize=4.5in
\epsffile{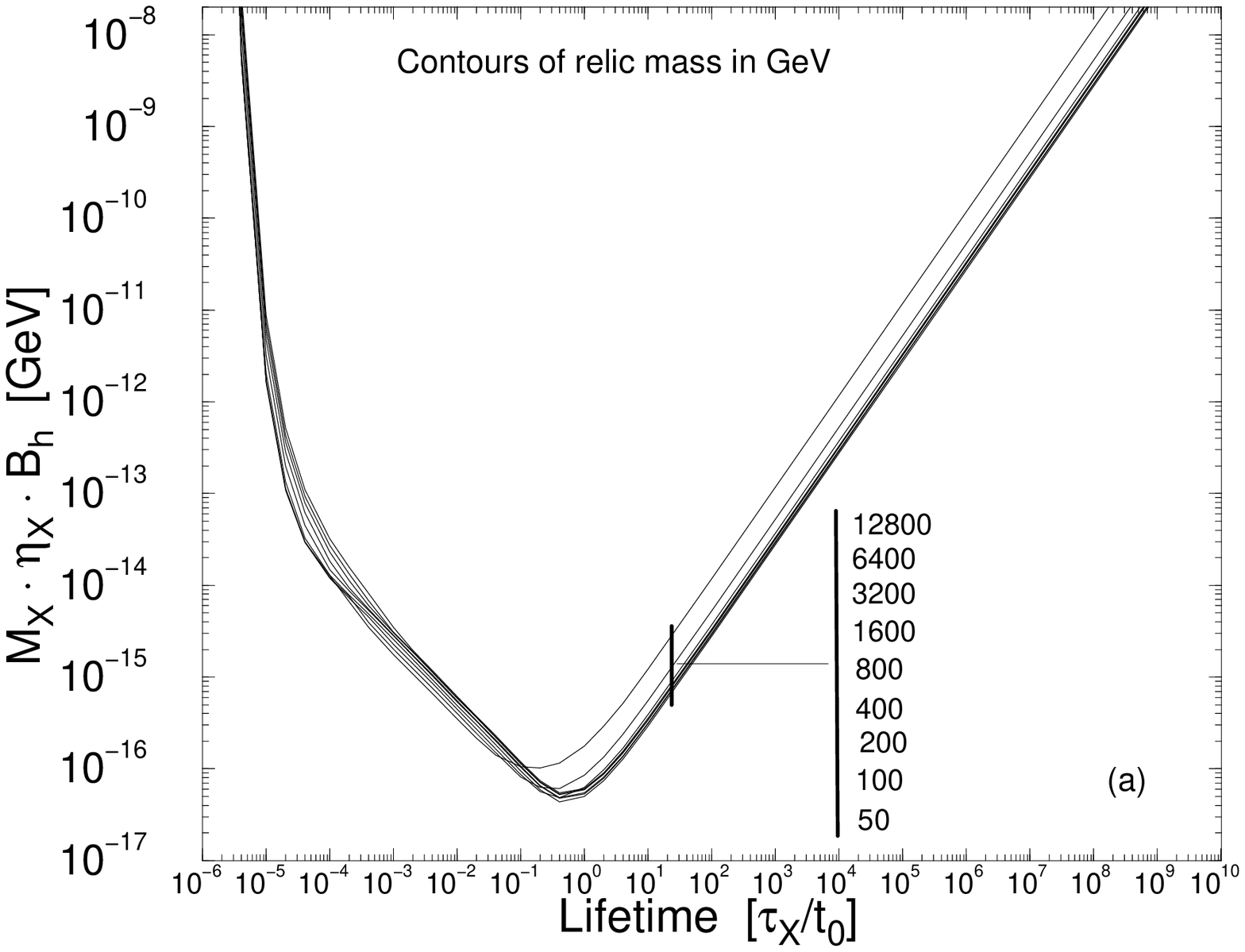}}
\centerline{
\epsfxsize=4.5in
\epsffile{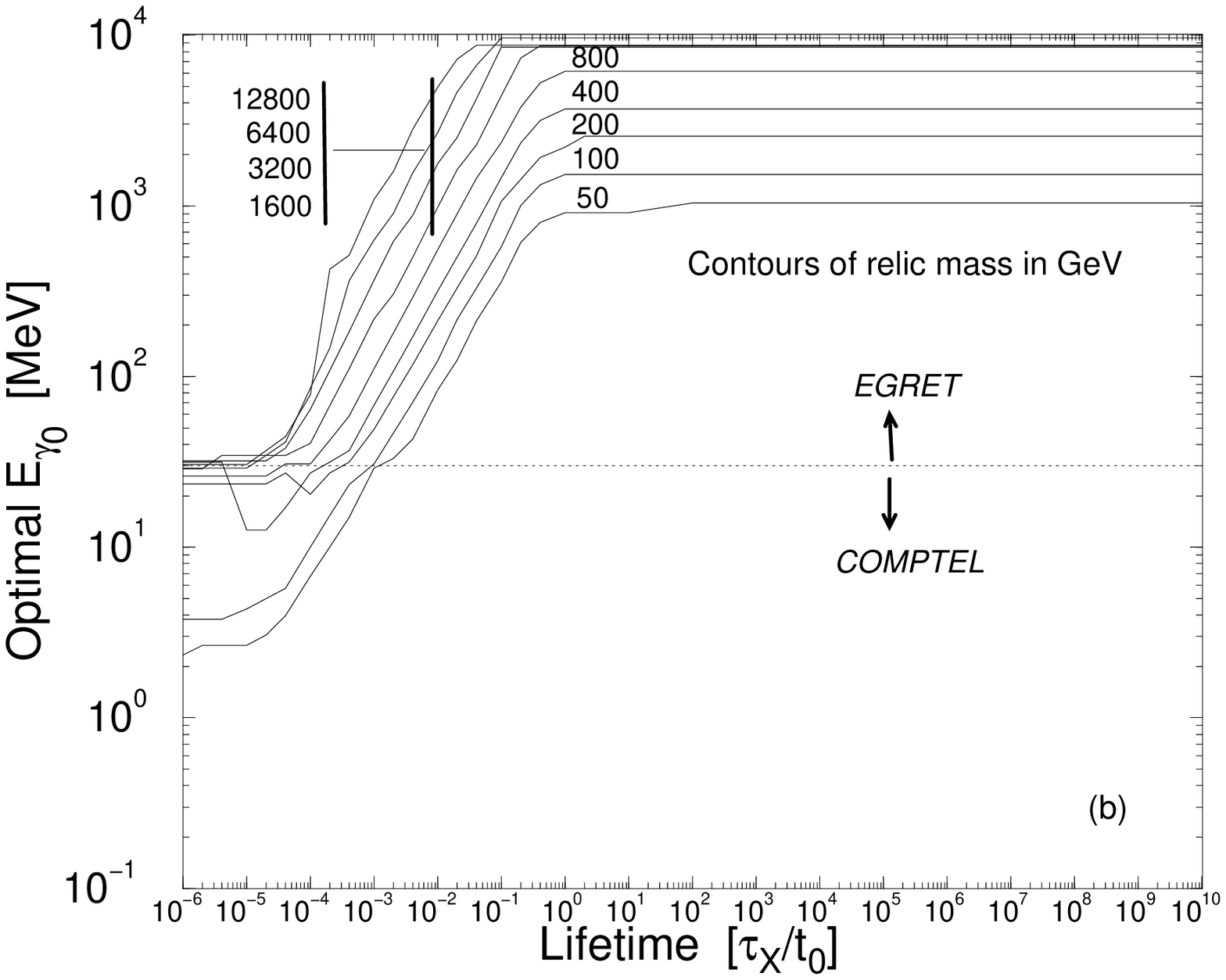}}
\caption{The upper graph shows the final relic density bound 
for 3-body hadronic decays with lifetimes in the indicated range.  
The bound scales linearly with the hadronic branching fraction 
of the relic $B_h$, although a branching ratio different from one 
does not strongly affect our bounds.  The upper limit on the relic 
density of $\sim 2 \times 10^{-8}$~GeV is roughly the critical density
corresponding to $\OmegaX h^2 \sim 1$.
The lower graph shows the optimal photon detection energy 
to obtain the best bound for a given lifetime.  
This graph is divided at $\Ephzero = 30$~MeV with a dotted line
to show which instrument provides the diffuse photon background bound
for a given lifetime.
}
\label{hadronic-bound-fig}
\end{figure}

By finding the maximum photon flux above background (i.e.\ one point
on each line in each graph of Fig.~\ref{hadronic-50-400-fig}), one
can derive the bound on the relic density for a given mass
lifetime as is done in Fig.~\ref{hadronic-bound-fig} (the
same procedure as in Fig.~\ref{2-body-bound-fig}).
We used the observational diffuse background fit as described 
in Sec.~\ref{measurements-sec}, and thus a bound for any given
mass and lifetime utilizes one (optimal) observational energy, as 
is shown in the lower graph of Fig.~\ref{hadronic-bound-fig}.
For example, for $\tauX/t_0 = 1$ one can see the trend in 
increasing $\Ephzero$ is to increase the photon flux 
(see Fig.~\ref{hadronic-50-400-fig}). Hence 
the best bound for this lifetime comes from 
observations of the most energetic photons.
On the other hand, for $\tauX/t_0 = 10^{-4}$, one finds that the
best bound for $\MX = 50$~GeV is roughly 
$\Ephzero \approx 7$~MeV\@.

\section{Implications of the Bounds}
\label{implications-sec}
\indent

There are two central results we can extract from
Figs.~\ref{2-body-bound-fig} and \ref{hadronic-bound-fig}
for both radiative and hadronic decays.  First,
a large range of lifetimes can be excluded for a relic with 
roughly the critical density.  Second, relics with densities 
considerably {\em below}\/ the critical density are excluded, 
which places a strong constraint on models with a long lived
massive particle.

\subsection{Relics with the critical density}
\indent

For relics with roughly the critical density that decay 
dominantly through a radiative channel, the bounds from the
diffuse background exclude lifetimes in the range
\begin{equation}
10^{12} \lsim \tauX \lsim 3 \times 10^{22} \;\> {\rm s}
\end{equation}
(using $t_0 = 10^{10}$~yr).  This bound applies to a relic
with {\em any}\/ mass $\MX \gsim 1$~MeV\@.  The upper bound
on the excluded lifetime of $3 \times 10^{22}$~s applies 
to the worst-case scenario with $\MX/2 = \Einj \approx 600$~GeV, where 
the upper bound increases to roughly $10^{27}$~s 
for $\MX/2 = \Einj = 25$~GeV and to roughly $10^{25}$~s 
for $\MX/2 = \Einj = 6400$~GeV\@.  The upper bounds increase
for larger mass relics $\MX \gsim 10$~TeV\@.
The upper bound on the excluded 
range is near $10^{28}$~s for small masses ${\cal O}(1 \; {\rm GeV})$.

For relics with roughly the critical density that decay
dominantly through hadronic channels, the bounds from
the diffuse background exclude lifetimes in the range
\begin{equation}
10^{12} \lsim \tauX \lsim 10^{26} \;\> {\rm s} 
\end{equation}
for masses $\MX = 50 \ra 10000$~GeV\@.  Unlike 2-body decays, 
the upper bound on the lifetime decreases as the mass of
the relic is increased.  This is because for masses 
beyond $10$~TeV, an increasing fraction
of the photons are scattered into lower energies where the 
diffuse photon bound is weaker.  For masses smaller than
$50$~GeV the upper bound on the lifetime is roughly similar,
but is absent once the mass is below the threshold for pion 
production.

\subsection{General bounds}
\indent

We have shown that a large range in relic lifetime can be 
excluded assuming the relic has roughly the critical density.
However, Figs.~\ref{2-body-bound-fig} and \ref{hadronic-bound-fig}
clearly show that the upper limit on the relic density is {\em much}\/
smaller that the critical density by several orders of magnitude.
In particular, models with a long lived particle that
does not have the critical density will be excluded if
the relic density exceeds our bounds.  
The translation of our bounds is relatively straightforward,
if most of the relics have not decayed prior to the earliest
redshift which we considered, $z_0 = 700$ (that is, if the lifetime
is longer than that in (\ref{shortest-lifetime-eq})).
Specifically,
\begin{equation}
\OmegaX h^2 \sim \frac{\MX \etaX}{2.5 \times 10^{-8} \; {\rm GeV}}
\end{equation}
for $\tauX/t_0 \gsim 10^{-4}$.

\section{Conclusions}
\label{conclusions-sec}
\indent

Utilizing the latest observations of the diffuse photon background 
found from EGRET and COMPTEL leads to the bounds summarized 
in Figs.~\ref{2-body-bound-fig} and \ref{hadronic-bound-fig}.  
Since the diffuse photon background is now 
well measured (up to $10$~GeV), these bounds establish a fixed 
upper limit on the relic density of long lived relics.  
We find that 3-body hadronic decays typically 
give better bounds than 2-body radiative decays for the 
same relic mass despite a larger total energy deposited 
into the final spectra for the latter decay mode.
The stronger bounds on hadronic decays are a direct result
of the low energy photons emitted from the fragmentation process
that produces pions which then decay into two photons.  However,
strong limits on radiative decays have also been obtained for
heavy mass relics since high energy photons are typically
scattered by either photon--photon scattering or pair
production.  In particular, relics with the critical
density and the masses considered here that
decay dominantly through radiative (hadronic) 
channels are excluded for lifetimes in the range
$10^{12} \lsim \tauX \lsim 10^{22} (10^{26})$~s.  
The upper bound on the excluded lifetime assumes
the worst-case, which is not necessarily the smallest
or largest mass.  For particular masses or lifetimes,
considerably more stringent bounds on the relic density 
can be read off from Figs.~\ref{2-body-bound-fig}, 
\ref{hadronic-bound-fig}.

The existence of strong bounds for both radiative and 
hadronic decays from the diffuse photon background 
provides a useful tool for those who consider long 
lived relics in particular particle physics or cosmological 
models.  If we assume that the relic has roughly the critical 
density, then we have seen that the lifetime must be far greater 
than the age of the universe. 

The bounds derived here allow for a more general analysis in that
we do not make any assumptions about the number density of the relic.
Thus, whether or not the relic is the dark matter, one is still able
to put strong constraints on the model.  Furthermore, it may be the
case that the model under consideration has more than one (meta-)stable
relic, one of which is a dark matter candidate. Such scenarios may 
arise in cases where there exist symmetries which are only broken by
higher dimensional operators. For instance, in many supersymmetric
models it is assumed that R parity is classically conserved.  In such cases 
one may expect that this symmetry will be broken by gravitational
effects leading to very long lived particles which may or may not be
dark matter candidates.  Bounds in such scenarios were discussed 
in Ref.~\cite{bere2} using limits from the positron flux in our
galaxy for the case of critical density\footnote{If the R symmetry 
is continuous, then there is a host of other constraints 
arising from Majoron production~\cite{roth1}.}.  These bounds are useful for
limiting the values of the couplings involved in the decay, which in general
are uncorrelated to all couplings that can be measured in accelerator
experiments.  However, if we do not make any assumptions regarding the
energy density of the relic then we may constrain couplings
that are accessible in collider experiments via a calculation of the
relic density.  The relic density may be determined by calculating the
annihilation cross section which will be a function of the accessible
parameters of the theory.  The drawback to this scenario is that one
must make some assumption about what is considered to be unnaturally small
for the symmetry breaking couplings.

\section*{Acknowledgments}
\indent

\begin{figure}
\centering
\epsfxsize=5.0in
\hspace*{0.0in}
\epsffile{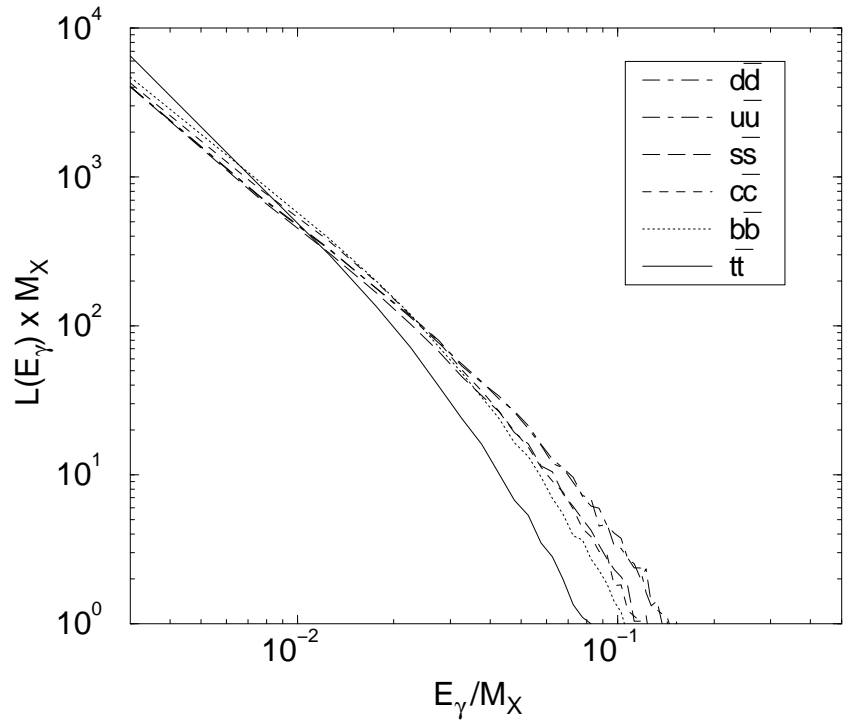}
\caption{The photon flux from selected final state quark pairs
from a 3-body decay of a relic.  Notice the photon spectrum for 
$t\overline{t}$ has an excess of very low energy $E_\ph/\MX \ll 0.01$ 
photons and a deficit of higher energy photons $E_\ph/\MX > 0.01$, 
compared with the photon spectra from lighter quark pairs.  
Note also that the spectra are effectively cutoff at 
$E_\ph = m_{\pi}/2$, although the mass used to generate 
these spectra $\protect\MX (= 800 \; {\rm GeV}) \gg m_{\pi}$,
and so the cutoff is not visible.
}
\label{compare-fig}
\end{figure}

We would like to thank C.~Fichtel for pointing us to recent 
EGRET data and D.~Gruber for discussions regarding the COMPTEL data.
We thank C.~Akerlof, G.~B.~Gelmini, G.~L.~Kane, E.~Nardi, D.~Seckel, 
and T.~Stanev for useful discussions.  We also thank M.~Birkel for 
enlightening correspondence related to Fig.~\ref{compare-fig}.  
I.~Z.~R. would like to thank the Aspen Center for Physics for 
its hospitality.  G.~D.~K. would like to thank G.~L.~Kane for 
encouragement and support.

\newpage

\newpage

\newpage 

\newpage

\newpage

\newpage

\newpage

\newpage

\newpage

\newpage

\newpage

\end{document}